\documentclass[desactivate]{aa}  

\usepackage{natbib}
\bibpunct{(}{)}{;}{a}{}{,}

\usepackage{graphicx}
\usepackage{revsymb}	
\usepackage{amsmath}
\usepackage{txfonts}
\usepackage{amssymb}
\usepackage{geometry} 
\usepackage[parfill]{parskip} 
\usepackage{slantsc}
\usepackage{array}
\usepackage{amsfonts}
	
\usepackage{epstopdf}	
\usepackage{epsfig}	

\usepackage{url}

\usepackage[colorlinks=true,linkcolor=black, urlcolor=black, citecolor=black]{hyperref}

\usepackage{color}
\usepackage[table]{xcolor}
\definecolor{lightgreen}{HTML}{90EE90}
\definecolor{salmon}{HTML}{FA9072}
\definecolor{cream}{HTML}{FFFFC2}

\usepackage{xfrac}

\usepackage{sidecap}
\usepackage[font=small, labelfont=bf]{caption}
\usepackage{subcaption}

\usepackage{threeparttable}

\begin{document}
   \title{Asteroseismic modelling strategies in the PLATO era}
      \subtitle{II. Automation of seismic inversions and quality assessment procedure}
\author{J. B\'{e}trisey\inst{1} \and G. Buldgen\inst{1,2} \and D. R. Reese\inst{3} \and G. Meynet\inst{1}}
\institute{Observatoire de Genève, Université de Genève, Chemin Pegasi 51, 1290 Versoix, Suisse\\email: 	\texttt{Jerome.Betrisey@unige.ch}
\and STAR Institute, University of Liège, 19C Allée du 6 Août, 4000 Liège, Belgium
\and  LESIA, Observatoire de Paris, Université PSL, CNRS, Sorbonne Université, Université Paris Cité, 5 place Jules Janssen, 92195 Meudon, France}
\date{\today}

\abstract{In the framework of the PLATO mission, to be launched in late 2026, seismic inversion techniques will play a key role in the mission precision requirements of the stellar mass, radius, and age. It is therefore relevant to discuss the challenges of the automation of seismic inversions, which were originally developed for individual modelling.}
{We tested the performance of our newly developed quality assessment procedure of seismic inversions, which was designed in the perspective of a pipeline implementation.}
{We applied our assessment procedure on a testing set composed of 26 reference models. We divided our testing set into two categories, calibrator targets whose inversion behaviour is well known from the literature and targets for which we assessed manually the quality of the inversion. We then compared the results of our assessment procedure with our expectations as a human modeller for three types of inversions, the mean density inversion, the acoustic radius inversion, and the central entropy inversion.}
{We found that our quality assessment procedure performs as well as a human modeller. The mean density inversion and the acoustic radius inversion are suited for a large-scale application, but not the central entropy inversion, at least in its current form.}
{Our assessment procedure showed promising results for a pipeline implementation. It is based on by-products of the inversion and therefore requires few numerical resources to assess quickly the quality of an inversion result.}


\maketitle

\section{Introduction}
Convective motions in the upper layers of solar-type stars generate a wide range of stellar oscillations. By studying these oscillations, asteroseismology enables us to probe the stellar interior and characterise the stellar parameters with a precision and accuracy that is difficult to match with other standard techniques for non-binary stars. Asteroseismology experienced a rapid development in the past two decades. The launch of space-based photometry missions such as CoRoT \citep{Baglin2009}, \emph{Kepler} \citep{Borucki2010}, and TESS \citep{Ricker2015} initiated the so-called photometry revolution. The unprecedented data quality from these missions allows us to use cutting edge techniques, the so-called seismic inversions \citep[see e.g.][]{Reese2012,Buldgen2015a,Buldgen2015b,Buldgen2018,Betrisey&Buldgen2022}, that were until then restricted to helioseismology \citep[see e.g.][for reviews]{Basu&Antia2008,Kosovichev2011,Buldgen2019e,JCD2021,Buldgen2022c}. Such seismic inversions were applied to various asteroseismic targets \citep[see e.g.][]{DiMauro2004,Buldgen2016b,Buldgen2016c,Buldgen2017c,Bellinger2017,Buldgen2019b,Bellinger2019d,Buldgen2019f,
Kosovichev&Kitiashvili2020,Salmon2021,Bellinger2021,Betrisey2022,Buldgen2022b,Betrisey2023_rot,Betrisey2023_AMS_surf}. In the near future, asteroseismic modelling will play a key role in the PLATO mission \citep{Rauer2014} for a determination of the stellar mass, radius, and age meeting the mission precision requirements (1-2\% in radius, 15\% in mass, and 10\% in age for a Sun-like star). It is therefore relevant to confront the current modelling strategies and discuss the remaining challenges for PLATO such as the choice of the physical ingredients \citep[see e.g.][]{Buldgen2019f,Betrisey2022}, the so-called surface effects \citep[see e.g.][]{Basu1996,Kjeldsen2008,Ball&Gizon2014,Sonoi2015,Ball&Gizon2017,Nsamba2018,Jorgensen2020,
Jorgensen2021,Cunha2021,Betrisey2023_AMS_surf} and stellar activity \citep[see e.g.][]{Broomhall2011,Santos2018,Santos2019_sig,Santos2019_rot,Howe2020,Thomas2021,Santos2021}.

In the first article of this series of papers, we presented a modelling strategy efficiently damping the surface effects and providing precise and accurate stellar parameters, based on the combination of a mean density inversion and a fit of frequency separation ratios \citep[][hereafter JB23]{Betrisey2023_AMS_surf}. Stellar seismic inversions were originally developed for solar modelling, and methods to assess the quality of the inversions were naturally investigated \citep[see e.g.][]{Pijpers&Thompson1992,Pijpers&Thompson1994,Rabello-Soares1999,Reese2012}. However, such methods cannot be applied in their current form to asteroseismic targets, because some of the simplifying hypotheses are only verified for the solar case, and the quality of an inversion is assessed manually by checking diagnostic plots and based on the experience of the modeller. The results of JB23 comforted us in the idea that a mean density inversion would be compatible with a pipeline approach. However, we also encountered a few lower quality inversion results that should be used with caution. In this study, we therefore propose a quality assessment procedure of seismic inversions that can be implemented in a pipeline. We considered three different types of inversion, the mean density inversion \citep{Reese2012}, the acoustic radius inversion \citep{Buldgen2015a}, and the central entropy inversion \citep{Buldgen2018}. We tested our assessment procedure on six calibrator models that are intensively studied targets in the literature: the Sun \citep[see e.g.][]{Reese2012}, Kepler-93 \citep[see e.g.][]{Betrisey2022}, 16 Cyg A and B \citep[see e.g.][]{Buldgen2016b,Buldgen2022b}, and $\alpha$ Cen A and B \citep[see e.g.][]{Reese2012,Salmon2021}. This procedure was then applied to 20 additional reference models for which we checked manually the diagnostic plots. 

In Sec. \ref{sec:modelling_strategy} we describe the different inversions that we investigated and our testing set. In Sec. \ref{sec:quality_assessment_procedure} we present our quality assessment procedure that is then applied on our testing set in Sec. \ref{sec:results}. We discuss in Sec. \ref{sec:discussion} best practices to consider for a large-scale application, and in Sec. \ref{sec:conclusions} we draw the conclusions of our study.

\section{Modelling strategy}
\label{sec:modelling_strategy}

\subsection{Seismic inversions}
In the paper, we use the following terminology. A seismic inversion takes as input a `reference' model, which is typically the optimal model from a local or global modelling strategy. In our study, most of the reference models come from a Markov chain Monte Carlo (MCMC) fitting the individual frequencies and the classical constraints (e.g. effective temperature, metallicity, luminosity), and for the $\alpha$ Cen binary system, a Levenberg Marquardt approach \citep[see e.g.][]{Frandsen2002,Teixeira2003,Miglio&Montalban2005} was employed. As a sidenote, we note that the interferometric radius can also serve as a classical constraint. However, except for specific cases \citep[e.g.][]{Pijpers2003,Huber2012,White2013}, such a measurement is rarely available. The inversion tries to recover the properties of an actual observed star or of a synthetic stellar model, which we call `target' or `observed' model. In our case, we considered both real observations, from the \emph{Kepler LEGACY} sample \citep{Lund2017} or from binary systems \citep{Kjeldsen2005,DeMeulenaer2010,Salmon2021,Buldgen2016b,Buldgen2022b}, and synthetic observations from \citet{Sonoi2015}, where the surface effects are emulated as realistically as possible with 3D hydrodynamic simulations of the upper stellar layers patched on a 1D structure. Based on the differences between the reference and observed frequencies, the inversion provides a small correction to a quantity of interest, in our case the mean density, the acoustic radius, and a central entropy indicator. We refer the quantity of interest including the small correction from the inversion as the `inverted' quantity.

In this section, we will provide a brief overview of the seismic inversion concepts that are pertinent to our study. We refer the reader to \citet{Gough&Thompson1991}, \citet{Gough1993}, \citet{Pijpers2006} and \citet{Buldgen2022c} for a more comprehensive discussion. The seismic inversions are based on the so-called structure inversion equation. By studying perturbations of the stellar oscillations at linear order, \citet{Lynden-Bell&Ostriker1967} and precursor studies \citep{Chandrasekhar1964,Chandrasekhar&Lebovitz1964,Clement1964} demonstrated that the equation of motion fulfils a variational principle. Using this finding for the individual frequencies, \citet{Dziembowski1990} showed that at first order, the frequency perturbation could be directly related to the structural perturbation through the structure inversion equation
\begin{equation}
\frac{\delta\nu^{n,l}}{\nu^{n,l}} = \int_{0}^{R} K_{a,b}^{n,l}\frac{\delta a}{a}dr + \int_{0}^{R} K_{b,a}^{n,l}\frac{\delta b}{b}dr + \mathcal{O}(\delta^2),
\label{eq:structure_inversion_equation}
\end{equation}
where $a$ and $b$ are two structural variables, $n$ is the radial order, $l$ is the harmonic degree, $\nu$ is the oscillation frequency, and $R$ is the stellar radius. $K_{a,b}^{n,l}$ and $K_{b,a}^{n,l}$ are the structural kernels, and the relative differences are computed with
\begin{equation}
\frac{\delta x}{x} = \frac{x_{\mathrm{obs}}-x_{\mathrm{ref}}}{x_{\mathrm{ref}}}.
\end{equation}
The indices `ref' and `obs' stand for reference and observed, respectively. We note that \citet{Dziembowski1990} originally derived Eq. \eqref{eq:structure_inversion_equation} for the $(\rho,c^2)$ structural pair, $\rho$ being the density and $c$ being the sound speed, and that the structure inversion equation can be adapted for any combination of physical variable that appears in adiabatic oscillations equations \citep[e.g.][]{Gough&Thompson1991,Gough1993,Elliott1996,Basu&JCD1997,Lin&Dappen2005,Kosovichev2011,Buldgen2017a,Buldgen2018}. Then, based on the relative differences between the observed and reference frequencies, the equations \eqref{eq:structure_inversion_equation} can be combined to compute a small correction to the reference model. Due to the limited number of modes in asteroseismology\footnote{about 50 in the best cases, compared to the few thousands for the Sun}, the goal is to define a quantity of interest, a so-called seismic indicator $t$, which concentrates all the information of the frequency spectrum. It typically takes the form
\begin{equation}
t = \int_0^R f(r)g(a)dr,
\end{equation}
where $f$ is a weight function function depending on the radius. The function $g$ is a function of the first structural variable $a$, that typically takes a simple form such as $g(a)=a$ or $g(a)=1/a$ \citep[see e.g.][for a review]{Buldgen2022c}. We note that a more general definition of the indicator can be used in specific cases \citep[see e.g.][]{Buldgen2015a}.

Compared to an approach solving directly the oscillations equation, like a MCMC would do, solving the structure inversion equation has the great advantage of not relying on the physics of the stellar evolution model. Indeed, the reference model is only a starting point for the inversion. In addition, the inversion also does not rely on the starting point. Using a different starting point in the parameter space, the inversion would still correct towards the exact value assuming that the starting point is in the linear regime, which means that Eq. \eqref{eq:structure_inversion_equation} is valid. The inversion can therefore provide a quasi-model independent correction. Several methods were developed to solve Eq. \eqref{eq:structure_inversion_equation}. Most of them rely on the optimally localised averages approach from \citet{Backus&Gilbert1968,Backus&Gilbert1970} or on the regularised least-squares technique from \citet{Tikhonov1963} \citep[see e.g.][]{Gough1985,JCD1990,Sekii1997,Buldgen2022c}. In our study, we used the subtractive optimally localised averages (SOLA) method \citep{Pijpers&Thompson1992,Pijpers&Thompson1994} which minimises the following cost function
\begin{align}
\mathcal{J}_{\bar{\rho}}(c_i) &= \int_0^1 \big(\mathcal{K}_{\mathrm{avg}} - \mathcal{T}_{t}\big)^2 dx 
                                        + \beta\int_0^1 \mathcal{K}_{\mathrm{cross}}^2dx + \lambda\left[k-\sum_i c_i\right] \nonumber\\
                                        &\quad+\tan\theta \frac{\sum_i (c_i\sigma_i)^2}{\langle\sigma^2\rangle} + \mathcal{F}_{\mathrm{Surf}}(\nu),
\end{align}
where $x=r/R$ and $k$ is a normalization constant which depends on the indicator's properties \citep[see e.g.][for a review]{Buldgen2022c}. The averaging $\mathcal{K}_{\mathrm{avg}}$ kernel and the cross-term kernel $\mathcal{K}_{\mathrm{cross}}$ are related to the structural kernels by
\begin{align}
\mathcal{K}_{\mathrm{avg}} &= \sum_i c_i K_{a,b}^{i}, \\
\mathcal{K}_{\mathrm{cross}} &= \sum_i c_i K_{b,a}^{i}.
\end{align}
The goal of the SOLA approach is to provide a good fit of the target function $\mathcal{T}_{t}$ while minimising the contribution of the cross-term and of the observational uncertainties. The variables $\beta$ and $\theta$ are trade-off parameters to adjust the balance between the different terms during the minimisation, and $\lambda$ is a Lagrange multiplier. The inversion coefficients are denoted by $c_i$, where $i\equiv(n,l)$ is the identification pair of an oscillation frequency, and $k$ is a normalisation constant. We defined $\langle\sigma^2\rangle = \sum_i^N\sigma_i^2$, where $\sigma_i$ is the $1\sigma$ uncertainty of the relative frequency difference and $N$ is the number of observed frequencies. The last term in the cost function, denoted by $\mathcal{F}_{\mathrm{Surf}}(\nu)$, is an empirical description of the surface effects. It introduces additional free parameters in the minimisation, in our case one, two or six depending on the surface effect prescription. These additional parameters come at the expense of the fit of the target function.

In this study, we considered three different indicators: $\bar{\rho},\tau,$ and $S_{\mathrm{core}}$. The indicator $\bar{\rho}$ is the mean density, and the target function of a mean density inversion is given by \citep{Reese2012}
\begin{align}
\mathcal{T}_{\bar{\rho}}(x) =  4\pi x^2 \frac{\rho}{\rho_R},
\end{align}
where $\rho_R=M/R^3$ and $M$ is the stellar mass. The trade-off parameters are fixed to $\beta=10^{-6}$ and $\theta=10^{-2}$, and we use the $(\rho,\Gamma_1)$ structural pair, where $\Gamma_1$ is the first adiabatic exponent.

The indicator $\tau$ is the acoustic radius
\begin{align}
\tau = \int_0^1 \frac{dx}{c},
\end{align}
and the target function of the inversion is given by \citep{Buldgen2015a}
\begin{align}
\mathcal{T}_{\tau}(x) = \frac{-1}{2\tau c}.
\end{align}
As for the mean density inversion, we use $\beta=10^{-6}$ and $\theta=10^{-2}$, and the $(\rho,\Gamma_1)$ structural pair.

The central entropy indicator $S_{\mathrm{core}}$ is defined as \citep{Buldgen2018}
\begin{align}
S_{\mathrm{core}} = \int_0^R \frac{f(r)}{S_{5/3}(r)}dr,
\end{align}
where $S_{5/3} =P/\rho^{5/3}$ is an entropy proxy and $P$ is the pressure. The weight function $f(r)$ is defined as follows
\begin{align}
f(r) &= \left[11r\exp\left(-29\left(\frac{r}{R}-0.12\right)^2\right) +  3r\exp\left(-2\left(\frac{r}{R}-0.14\right)^2\right)\right. \nonumber\\
       &\quad\left. + \frac{0.4}{1+\exp\left(\frac{1}{1.2}\left(\frac{r}{R}-1.7\right)\right)}\right]\cdot \tanh\left(50\left(1-\frac{r}{R}\right)\right),
\end{align}
This complicated weight function is designed to probe the core regions of the entropy proxy profile, while minimizing the upper layers where $S_{5/3}$ follows a plateau in the outer convective zone and takes on high values close to the outer boundary of the model. Therefore, this region must be efficiently damped in the cost function. The target function is then given by
\begin{align}
\mathcal{T}_{S_{\mathrm{core}}}(r) = \frac{-f(r)}{S_{\mathrm{core}}\cdot S_{5/3}(r)}.
\end{align}
This inversion is based the $(S_{5/3},Y)$ structural pair, where $Y$ is the helium mass fraction, and we use $\beta=\theta=10^{-4}$.

\subsection{Testing set}
\begin{table*}[t!]
\centering
\caption{Observational constraints of the targets from our testing set.}
\begin{tabular}{lcccccc}
\hline 
Nickname & KIC & $T_{\mathrm{eff}}$ & [Fe/H] & $L$  & $\nu_{max}$ & References \\ 
 &  & (K) & (dex) & $(L_\odot)$  & ($\mu$Hz) & \\ 
\hline \hline 
Model A &  & $5775.1\pm 90$ & $0.027\pm 0.10$ & $1.003\pm 0.191$ &  & 1 \\ 
Model B &  & $6725.8\pm 100$ & $0.027\pm 0.10$ & $3.862\pm 0.425$ &  & 1 \\ 
Model C &  & $6485.8\pm 100$ & $0.027\pm 0.10$ & $6.399\pm 0.704$ &  & 1 \\ 
Model D &  & $6431.9\pm 100$ & $0.027\pm 0.10$ & $2.969\pm 0.564$ &  & 1 \\ 
Model E &  & $6227.0\pm 100$ & $0.027\pm 0.10$ & $5.075\pm 0.558$ &  & 1 \\ 
Model F &  & $6103.3\pm 100$ & $0.027\pm 0.10$ & $2.137\pm 0.406$ &  & 1 \\ 
Model G &  & $5861.0\pm 90$ & $0.027\pm 0.10$ & $0.990\pm 0.188$ &  & 1 \\ 
Sun &  & $5772 \pm 85$ & $0.00 \pm 0.10$ & $1.00 \pm 0.03$ &  & 2 \\ 
$\alpha$ Cen A &  & $5795\pm 19$ & $0.24 \pm 0.01$ & $1.521\pm 0.015$ &  & 3 \\ 
$\alpha$ Cen B &  & $5231\pm 21$ & $0.24 \pm 0.01$ & $0.503\pm 0.007 $ &  & 3 \\ 
Kepler-93 & 3544595 & $5718 \pm 100$ & $-0.18 \pm 0.10$ & $0.82 \pm 0.03$ &  & 4 \\ 
Nunny & 6116048 & $6033 \pm 100$ & $-0.23 \pm 0.10$ & $1.85 \pm 0.07$ & $2126.9 \pm 5.3$ & 4 \\ 
Saxo2 & 6225718 & $6203 \pm 100$ & $-0.17 \pm 0.10$ & $2.13 \pm 0.08$ & $2364.2 \pm 4.8$ & 5 \\ 
Baloo & 6508366 & $6331 \pm 100$ & $-0.05 \pm 0.10$ & $6.78\pm 0.27$ & $958.3\pm  4.1$ & 5 \\ 
Doris & 8006161 & $5488 \pm 100$ & $0.34 \pm 0.10$ & $0.69 \pm 0.03$ & $3574.7 \pm 11.0$ & 4 \\ 
Arthur & 8379927 & $6067 \pm 150$ & $-0.10 \pm 0.15$ & - & $2795.3 \pm 6.0$ & 6 \\ 
 Carlsberg & 9139151 &$6043 \pm 100$ & $0.05 \pm 0.10$ & $1.60 \pm 0.06$ & $2690.4 \pm 11.8$ & 5 \\ 
Punto & 9139163 & $6400 \pm 60$ & $0.15 \pm 0.10$ & $3.68\pm 0.13$ & $1729.8\pm  6.1$ & 5 \\ 
Pinocha & 10454113 & $6177 \pm 100$ & $-0.07 \pm 0.10$ & - & $2357.2 \pm 8.7$ & 4 \\ 
Tinky & 11253226 & $6642 \pm 100$ & $-0.08 \pm 0.10$ & $4.44\pm 0.16$ & $1590.6 \pm 8.7$ & 4 \\ 
Dushera & 12009504 & $6179 \pm 100$ & $-0.08 \pm 0.10$ & $2.70 \pm 0.11$ & $1865.6 \pm 7.0$ & 4 \\ 
16 Cyg A & 12069424 & $5839 \pm 42$ & $0.096 \pm 0.026$ & $1.56 \pm 0.02$ &  & 7 \\ 
16 Cyg B & 12069449 & $5809 \pm 39$ & $0.052 \pm 0.021$ & $1.27 \pm 0.02$ &  & 7 \\ 
Barney & 12258514 & $5964 \pm 60$ & $0.00 \pm 0.10$ & $2.95 \pm 0.11$ & $1512.7 \pm 3.1$ & 4 \\ 
\hline 
\end{tabular}
{\par\small\justify\textbf{Notes.} (1) \citet{Sonoi2015}; (2) frequencies from measurement n°$01$ of \citet{Salabert2015} and $T_{\mathrm{eff}}$ from \citet{IAU2015-ResolutionB3}, the uncertainty of the classical constraints is adapted to match the expectation from a \emph{Kepler} observation; (3) see \citet{Salmon2021} and references therein; (4) \citet{Lund2017}; (5) frequencies and $\nu_{max}$ from \citet{Lund2017}, $T_{\mathrm{eff}}$ and [Fe/H] from \citet{Furlan2018}; (6) frequencies from \citet{Roxburgh2017}, $T_{\mathrm{eff}}$, [Fe/H], and $\nu_{max}$ from \citet{Lund2017}; (7) $T_{\mathrm{eff}}$ from \citet{White2013}, [Fe/H] from \citet{Ramirez2009}, and $L$ from \citet{Metcalfe2012}. Models A to G are synthetic models whose classical constraints are known exactly. The uncertainty was chosen to match the expectation from a \emph{Kepler} observation. The other targets are actual observations for which the luminosity was estimated using the spectroscopic parameters and Eq. (13) in JB23. The RUWE indicator of Gaia flags the parallax measurement of Arthur as unreliable, and the $K_s$-magnitude measurement of Pinocha is unreliable as well. \par}
\label{tab:observational_constraints}
\end{table*}

Our testing set is composed of 26 reference models that we divided in two categories. The first category is composed of six calibrator targets. For these calibrators, advanced and extensive modelling were conducted in the literature. The behaviour of the seismic inversions that were carried out on these targets was therefore thoroughly investigated. We considered the following calibrator targets: the Sun \citep[see e.g.][]{Reese2012}, Kepler-93 \citep[see e.g.][]{Betrisey2022}, 16 Cyg A and B \citep[see e.g.][]{Buldgen2016b,Buldgen2022b}, and $\alpha$ Cen A and B \citep[see e.g.][]{Reese2012,Salmon2021}. The second category is composed of 18 targets that we selected either from the \emph{Kepler LEGACY} sample \citep{Lund2017} or from \citet{Sonoi2015}. These targets were less studied by the literature than the targets from the first category and they cannot be considered as calibrators. To assess the inversion quality of these targets independently from the quality assessment procedure of Sec. \ref{sec:quality_assessment_procedure}, we checked manually how well the target function is reproduced by the averaging kernel. We note that this check allows us to discard robustly the most problematic inversion results, but that there is a grey zone that depends on the experience of the modeller and where it is unclear whether the inversion result is robust. This uncertainty can be lifted by conducting a more extensive analysis, namely by generating a set of models representative of the target and investigating the behaviour of the inversion on the set, as it was done for the calibrator targets \citep[see e.g.][]{Betrisey2022,Buldgen2022b}. The second category is however relevant in the sense that we can check whether the automatic assessment procedure of Sec. \ref{sec:quality_assessment_procedure} performs equivalently to a human modeller.

We summarised in Table \ref{tab:observational_constraints} the observational constraints of our testing set. The reference model of Kepler-93 is the Model$_1$ from \citet{Betrisey2022}. For $\alpha$ Cen A and B, we considered two sets of reference models, including overshooting in $\alpha$ Cen A or not \citep[see Table 4 in][]{Salmon2021}. We note that the reference models of $\alpha$ Cen B are different, because \citet{Salmon2021} evolved both stars of the binary system simultaneously in the minimisation. For the rest of the targets, the reference models were obtained with a MCMC fitting the individual frequencies and the classical constraints. The detailed modelling procedure is described in JB23, as well as the grid of models used for the MCMC. The solar model, model G, Baloo, Punto, and Tinky were added for this study and we proceeded exactly as for the targets from JB23. For the Sun, we selected the frequencies of the measurement n°01 from \citet{Salabert2015} and the effective temperature from \citet{IAU2015-ResolutionB3}. The observational uncertainties of the classical constraints were adapted to match the precision of a target observed by \emph{Kepler}: $85$~K for the effective temperature, $0.1$ dex for the metallicity, and $0.03L_\odot$ for the luminosity. The data of model G was taken from \citet{Sonoi2015}, and the data of Baloo, Punto, and Tinky from \citet{Lund2017}.

\section{Quality assessment procedure}
\label{sec:quality_assessment_procedure}
Before we introduce the assessment procedure, we would like to clarify some terminological aspects. Synthetic models with known structures have been extensively employed to validate and establish the reliability of inversions \citep[in particular in][for the inversions of this study]{Reese2012,Buldgen2015a,Buldgen2018}. In a concrete application on observed data, it is not possible to verify the accuracy of an inversion. However, it is essential to verify the numerical stability of inversions, which can be compromised by factors such as data quality or unaccounted nonlinearities \citep[as observed in the case of $\alpha$ Cen A;][]{Salmon2021}. Numerical instability can indeed jeopardize the reliability of the inversion results. Previously, manual scrutiny of diagnostic plots was the norm for assessing stability, but this article introduces an automated procedure for this purpose. Therefore, when we label an inversion as stable or successful, it indicates that the inversion was numerically stable. Conversely, if an inversion is labelled as a failure, it means that it was numerically unstable. In this case, the inversion result should be treated with caution.

Our quality assessment procedure is based on two tests, which were specially designed to be compatible with a pipeline and replace the manual verifications that were until then required to assess the quality of an inversion. These tests are based on so-called quantifiers, whose value corresponds to three different flags: \textit{reject} the inversion result, \textit{check} manually the inversion result by generating a set of models representative of the target and study the behaviour of the inversion on the set \citep[see e.g.][]{Betrisey2022,Buldgen2022b}, and \textit{accept} the inversion result. The first test measures the quality of the fit of the target function by the averaging kernel. We call `K-flag' the outcome of this first test. The second test quantifies the randomness of the inversion coefficients. Indeed, we noted in JB23 that smooth structures appear in successful inversions. If the inversion gets unstable,  these structures break down, and the inversion coefficients tend to be randomly distributed. We call `R-flag' the outcome of this second test. We recommend to use our assessment procedure as follows; the K-flag is first computed, and the R-flag is then evaluated
only if the inversion was not rejected by the K-flag. Indeed, the goal of the first test is only to remove the inversion results that are clearly wrong prior to the second test, which is the core of our assessment procedure.

\subsection{K-flag}
\begin{figure}
\centering
\begin{subfigure}[b]{.43\textwidth}
  \includegraphics[width=.99\textwidth]{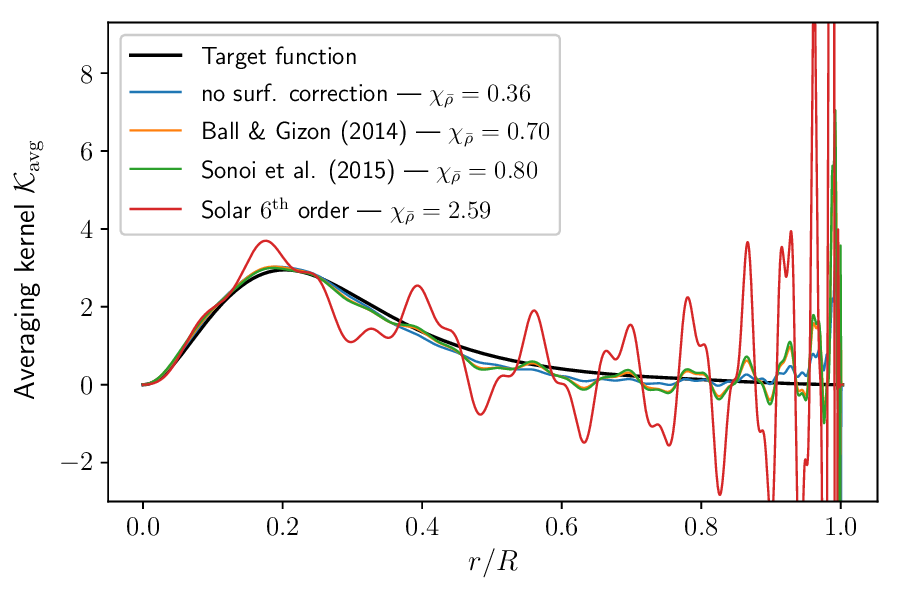}  
  \label{fig:Sun01_averaging_kernels}
\end{subfigure}
\begin{subfigure}[b]{.43\textwidth}
  \includegraphics[width=.99\textwidth]{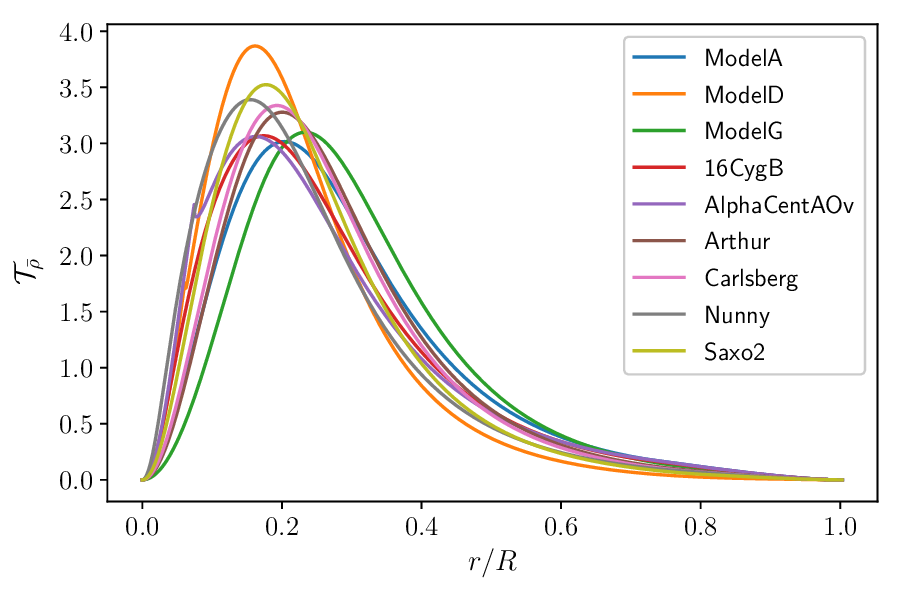}  
  \label{fig:variation_Tfunc_rho}
\end{subfigure} 
\caption{Averaging kernels of the solar model and variation of the target function for a selection of models from our testing set. \textit{Top panel:} Averaging kernels of the solar model by considering different surface effect prescriptions. \textit{Bottom panel:} Variation of the target function of a mean density inversion for a selection of models from our testing set.}
\label{fig:K_flag_target_functions}
\end{figure}

The K-flag assesses the quality of the fit of the target function by the averaging kernel, and is a binary flag which takes following values: \textit{accept} or \textit{reject}. The quality of the fit of the target function is an important aspect of a seismic inversion because a poor fit may induce a non-physical inversion result. The question of the quality of the fit of the target function was raised at the same time as the seismic inversions were developed, and it was proposed to compute the square of the $L_2$-norm of the difference between the averaging kernel and the target function \citep[see e.g.][]{Pijpers&Thompson1992,Pijpers&Thompson1994,Rabello-Soares1999,Reese2012}
\begin{align}
\chi_t &= ||\mathcal{K}_{\mathrm{avg}}-\mathcal{T}_t||^2_2, \\
           &= \int_0^1 \left(\mathcal{K}_{\mathrm{avg}}(x)-\mathcal{T}_t(x)\right)^2 dx.
\label{eq:chi_t_avg}
\end{align}
This quantifier was however introduced for solar inversions and implicitly assumes that the target function of the different reference models has a comparable amplitude, which is valid in solar modelling. Additionally, we point out that we are working with a scaled radius in the formulation of the kernels so that the domain of the kernels in all cases is $[0, 1]$ and also, that the averaging kernels are always normalised to have an integral of 1 over this domain. It is therefore possible to compare the inversions by looking at the absolute value of $\chi_t$. In the top panel of Fig. \ref{fig:K_flag_target_functions}, we illustrate the averaging kernels of the solar model by considering several prescriptions for the surface effects. In this conditions, the target function do not change. In the context of a space-based photometry missions such as \emph{Kepler} or PLATO, the solar-type stars which are observed cover a mass range between $0.8M_\odot$ and $1.6M_\odot$. The amplitude of the target functions varies significantly between the different targets, as shown in the bottom panel of Fig. \ref{fig:K_flag_target_functions}, and it is less meaningful to compare them directly with $\chi_t$. However, $\chi_t$ can still be used to filter the most problematic inversion results. Indeed, if the averaging kernel is unable to reproduce the target function (see examples in Fig. \ref{fig:Kflag_abberant_cases}), $\chi_t$ takes a large value. By defining a rejection threshold large enough not to be sensitive to the specific amplitude of the target function, outlying inversion results with an extreme value of $\chi_t$ can still be sorted out. We note that this threshold should not be interpreted as an exact threshold because of the limitations that we mentioned earlier, but rather as a filter in preparation for the second test. Based on our testing set of main-sequence solar-type stars, we defined in Table~\ref{tab:K_flag_rejection_criteria} a rejection threshold for each of the inversions considered in this study. The form of the target function is specific to each type of inversion. The rejection threshold therefore depends on the type of inversion, but it is always possible to identify such threshold.

For the reasons given above, we have opted for a pragmatic way of determining the rejection threshold based on our testing set. From a theoretical standpoint however, it would be possible to obtain a more objective estimate of this threshold by considering the following idea. Let us denote the $\chi_t$ obtained using Eq. \eqref{eq:chi_t_avg} as $\chi_t^{\mathrm{avg}}$. We construct a substantial number of pairs of models that we are able to distinguish asteroseismically (for example by looking on the edges of uncertainty boxes in HR-like diagrams). The models in these pairs have target functions $\mathcal{T}_t^j$ and $\mathcal{T}_t^k$, respectively. Then, we calculate $\chi_t$ using the difference between those two target functions and take the supremum
\begin{align}
\chi_t^{\mathrm{sup}} = \sup_{j,k}\left\lbrace\int_0^1 \left(\mathcal{T}_t^j(x)-\mathcal{T}_t^k(x)\right)^2 dx\right\rbrace.
\end{align}
Assuming that we have a reference model with $\chi_t^{\mathrm{avg}} > \chi_t^{\mathrm{sup}}$, it would imply that the averaging kernel of this reference model fits the target function less efficiently than a model that can be rejected based on the asteroseismic constraints alone. To generate the substantial number of model pairs, we could use the MCMC steps which are on the edges of uncertainty boxes. In practice however, the current version of the MCMC interpolates within the parameter space, but does not provide an interpolated structure. Accurately interpolating this structure would be quite challenging and could lead to a notable slowdown in the minimisation process, which is already quite expensive. Another option would be to use the grid models that are on the border of a $1\sigma$ or $2\sigma$ uncertainty ball around the MCMC solution. Further investigations are needed to determine the level of grid density required for generating of a sufficient amount of model pairs. Additionally, we anticipate challenges with the grid model structures from actual missions such as PLATO. Indeed, the grids utilized for these missions covers the entire parameter space of interest, taking a very large amount of storage space. Therefore, only reduced or minimal structures are saved and additional computations are needed to restore complete structures. In any case, the determination of $\chi_t^{\mathrm{sup}}$ is probably too expensive to be employed on each an every target in a pipeline, but it may be useful to apply this procedure to benchmarks in the future to improve the estimate of the rejection thresholds adopted in this study.

\begin{table}
\centering
\caption{Rejection threshold of the K-flag.}
\begin{tabular}{lc}
\hline 
 & Rejection criterion \\ 
\hline\hline 
$\bar{\rho}$ inversion & $\chi_{\bar{\rho}} > 4$ \\ 
$\tau$ inversion & $\chi_{\tau} > 2$ \\ 
$S_{\mathrm{core}}$ inversion & $\chi_{S_{core}} > 1$ \\ 
\hline 
\end{tabular} 
\label{tab:K_flag_rejection_criteria}
\end{table}


\subsection{R-flag}
\begin{figure*}[htp!]
  \centering
\begin{subfigure}[b]{.33\textwidth}
  \includegraphics[width=.99\linewidth]{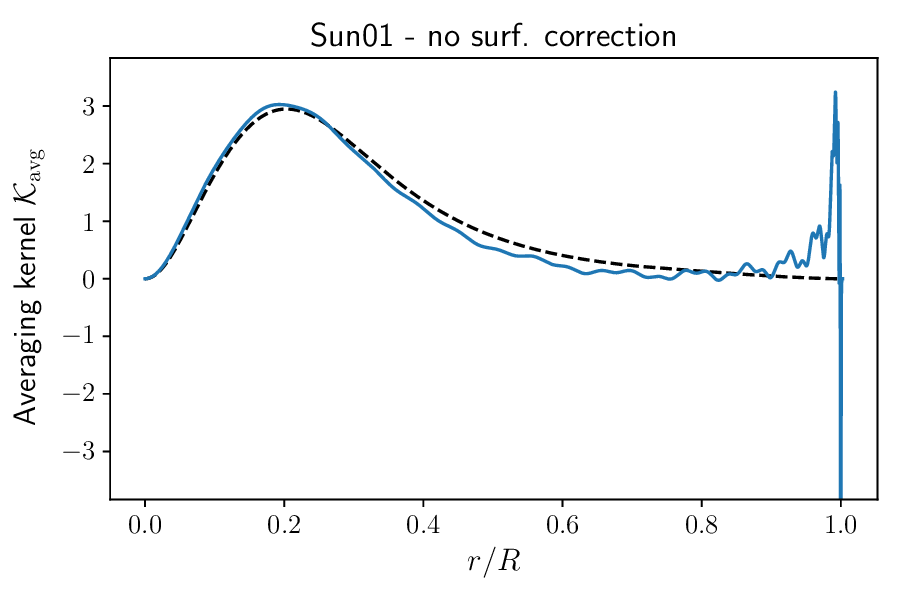}  
\end{subfigure}
\begin{subfigure}[b]{.33\textwidth}
  \includegraphics[width=.99\linewidth]{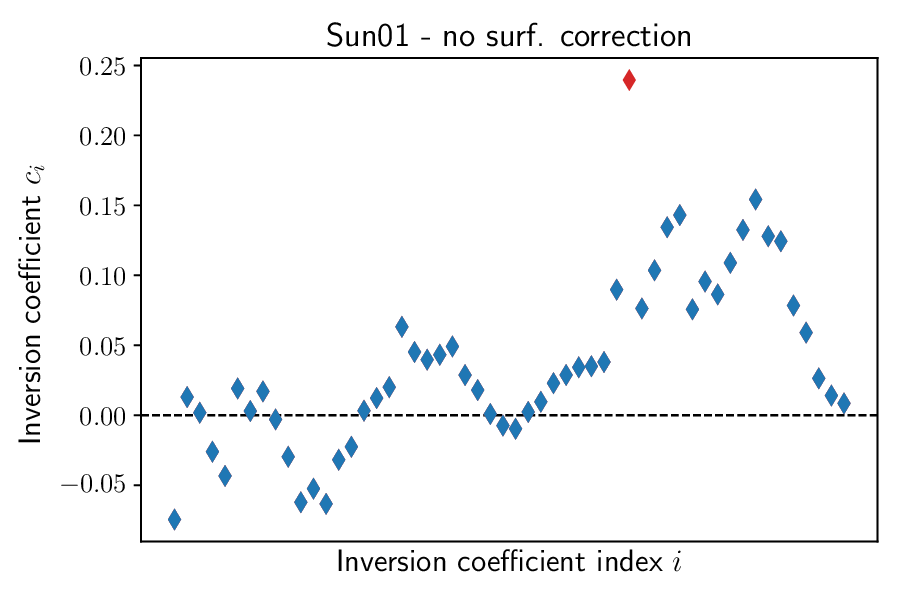} 
\end{subfigure}
\begin{subfigure}[b]{.33\textwidth}
  \includegraphics[width=.99\linewidth]{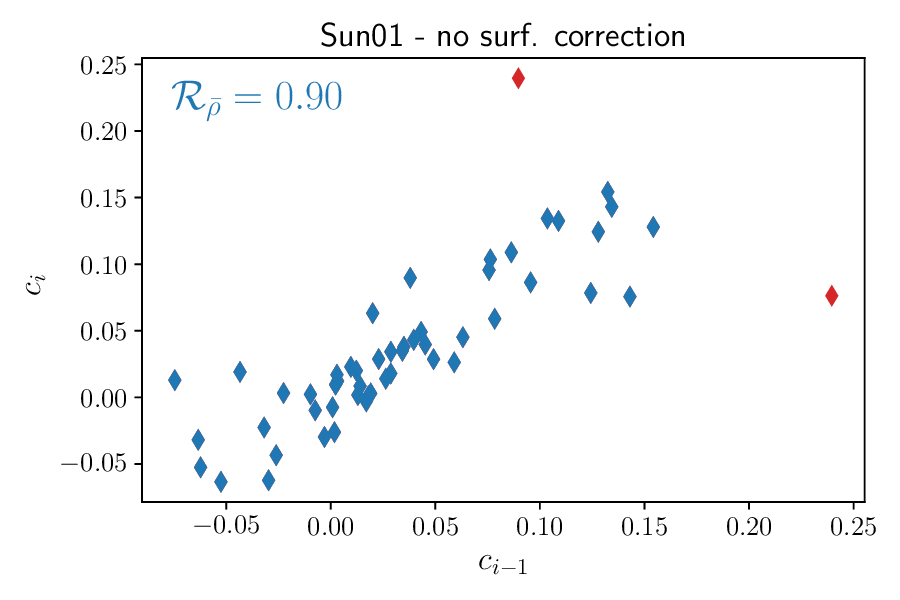} 
\end{subfigure}
\begin{subfigure}[b]{.33\textwidth}
  \includegraphics[width=.99\linewidth]{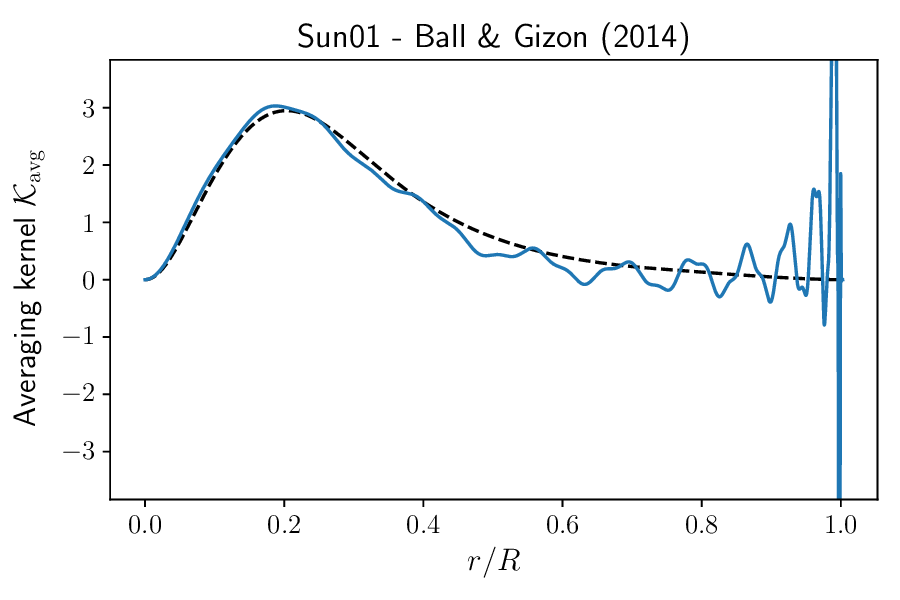} 
\end{subfigure}
\begin{subfigure}[b]{.33\textwidth}
  \includegraphics[width=.99\linewidth]{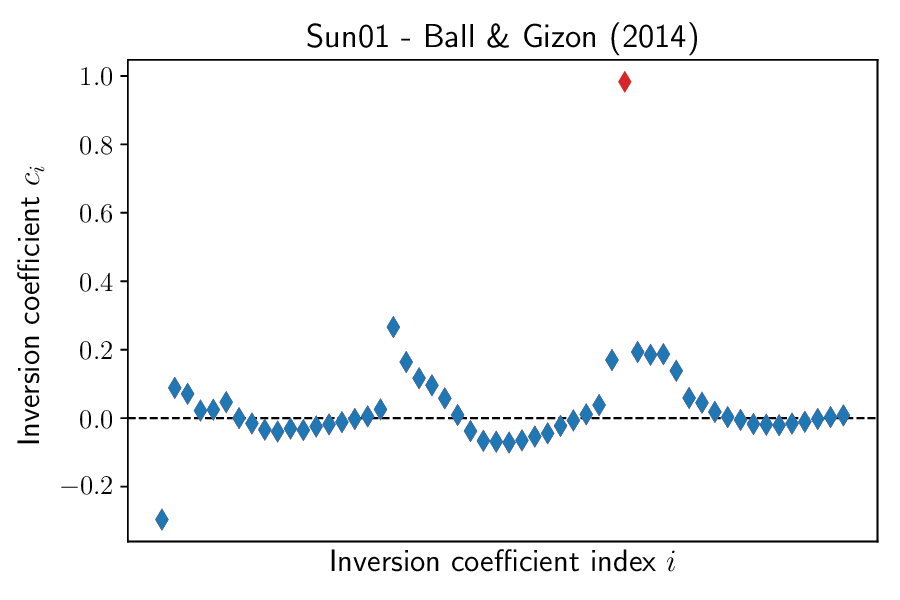}  
\end{subfigure}
\begin{subfigure}[b]{.33\textwidth}
  \includegraphics[width=.99\linewidth]{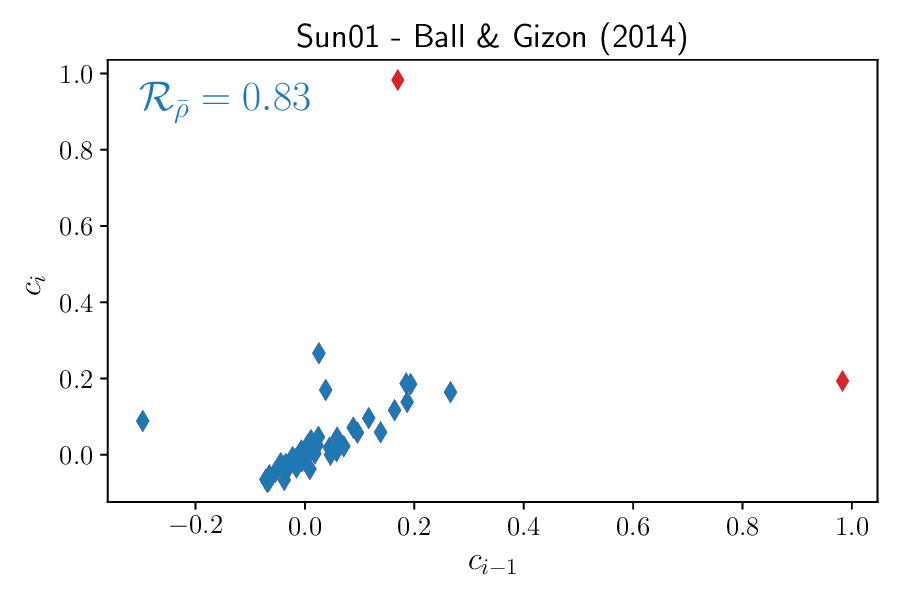} 
\end{subfigure}
\begin{subfigure}[b]{.33\textwidth}
  \includegraphics[width=.99\linewidth]{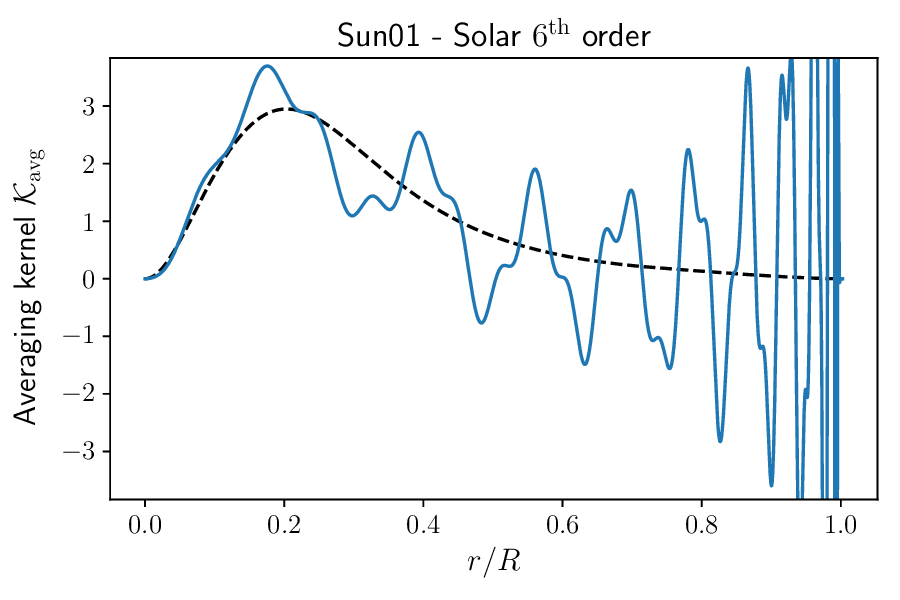}  
\end{subfigure}
\begin{subfigure}[b]{.33\textwidth}
  \includegraphics[width=.99\linewidth]{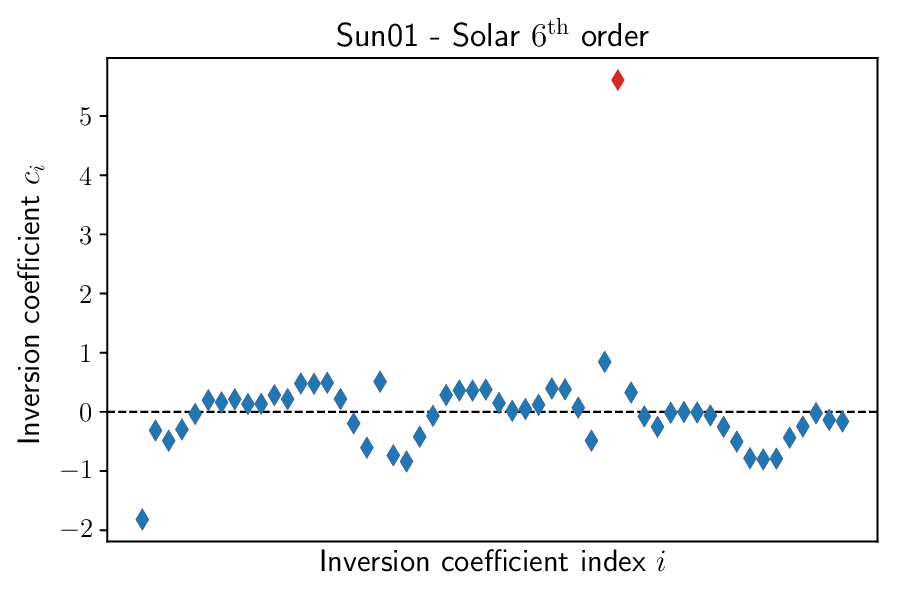}  
\end{subfigure}
\begin{subfigure}[b]{.33\textwidth}
  \includegraphics[width=.99\linewidth]{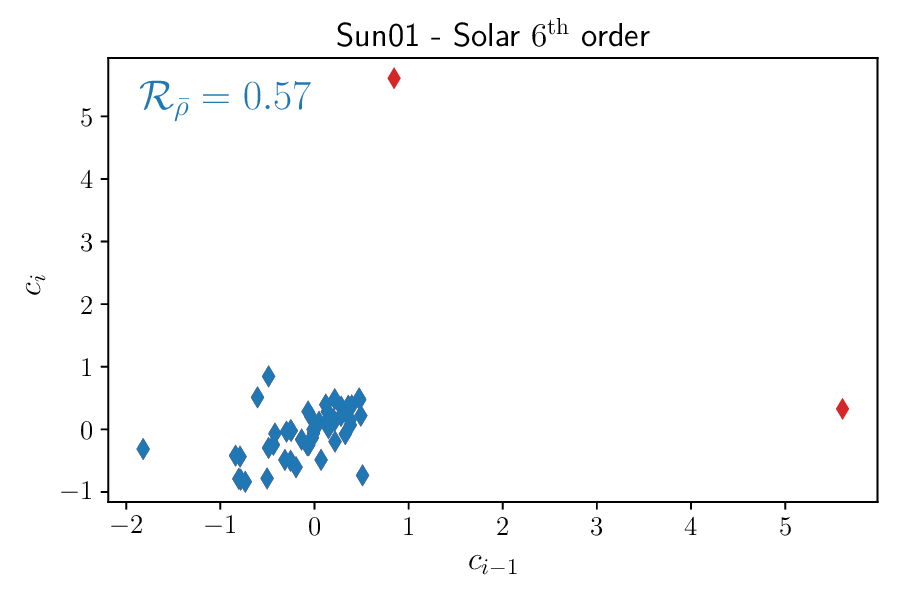} 
\end{subfigure}
\begin{subfigure}[b]{.33\textwidth}
  \includegraphics[width=.99\linewidth]{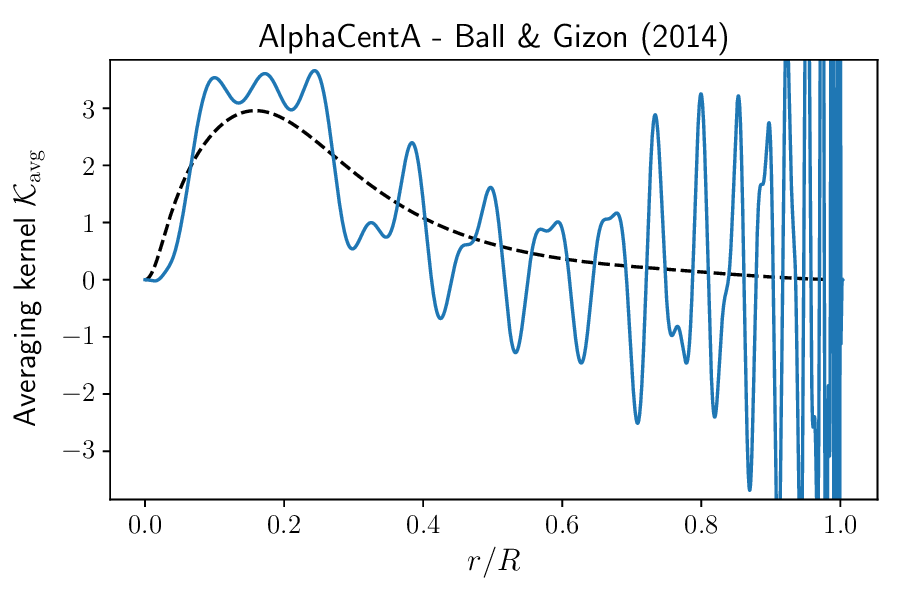}  
\end{subfigure}
\begin{subfigure}[b]{.33\textwidth}
  \includegraphics[width=.99\linewidth]{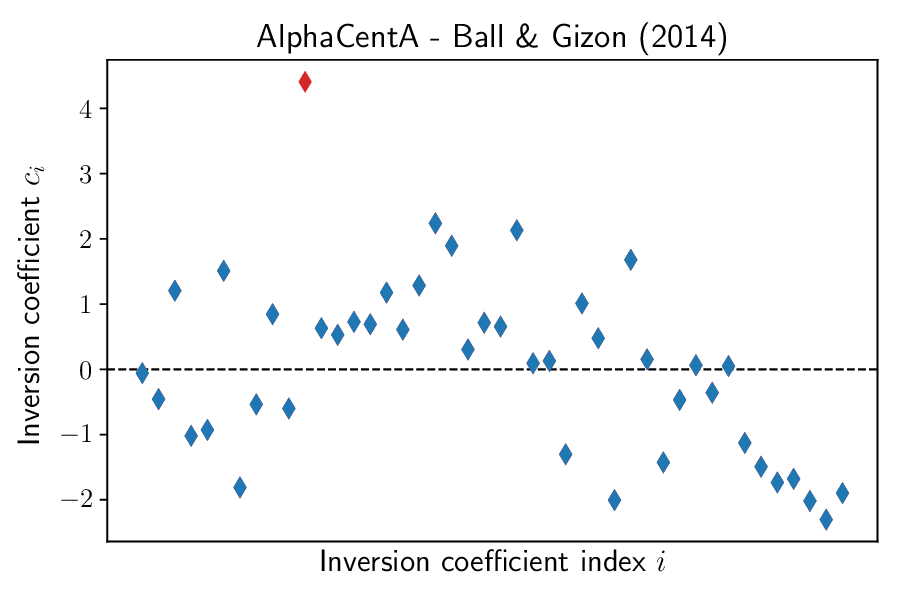}  
\end{subfigure}
\begin{subfigure}[b]{.33\textwidth}
  \includegraphics[width=.99\linewidth]{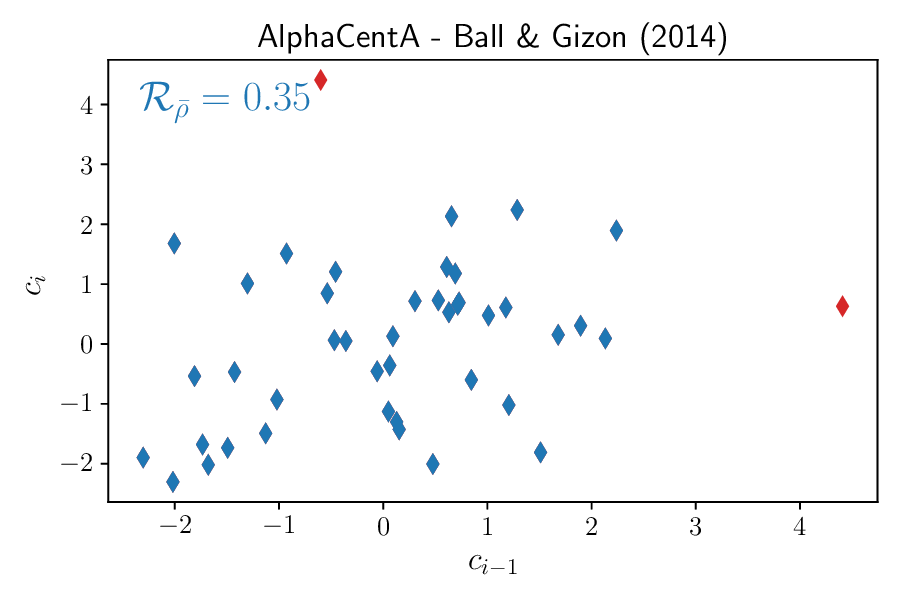} 
\end{subfigure}
\caption{Diagnostic plots of the solar model and of the $\alpha$ Cen A model. \textit{From top to bottom:} Diagnostic plots of the solar model by neglecting the surface effects, by using the \citet{Ball&Gizon2014} surface effect prescription, and by using a sixth order polynomial for the surface effects, and of $\alpha$ Cen A by using the \citet{Ball&Gizon2014} prescription. \textit{Left column:} Fit of the target function by the averaging kernel. \textit{Central column:} Inversion coefficients. \textit{Right column}: Lag plot of the inversion coefficients. The points in red are the values that were excluded.}
\label{fig:lag_examples}
\end{figure*}

In a seismic inversion, we assume that the relative frequency differences are independent measurements, but under simplifying hypotheses, one can show that the acoustic frequencies follow an asymptotic relation \citep{Shibahashi1979,Tassoul1980}:
\begin{align}
\nu_{n,l} = \left(n + \frac{l}{2} + \epsilon\right)\Delta\nu + \mathcal{O}(\Delta\nu^2),
\end{align}
where $\epsilon$ is a phase and $\Delta\nu$ is the large separation. In our previous study (Appendix A of JB23), we noted that the inversion coefficients of a stable inversion tend to show smooth structures. Because of the asymptotic behaviour of the frequencies, the same seismic information can be shared by multiple frequencies and it is therefore not unexpected to find smooth structures in the inversion coefficients, as illustrated in Fig. \ref{fig:lag_examples} for the solar model. If the target function is less well reproduced by the averaging kernel, these smooth structures break down and the inversion coefficients appear to be more randomly distributed, as illustrated in Fig. \ref{fig:lag_examples} for $\alpha$ Cen A. In JB23, we proposed to quantify this observation by looking at the lag plot \citep[see e.g.][for a reference handbook]{Heckert2002} of the inversion coefficients. Indeed, as shown in the right column of Fig. \ref{fig:lag_examples}, the inversion coefficients of a stable inversion tend to be positively correlated in the lag plot, assuming a lag of one. In our previous study, we suggested to quantify this correlation with the Pearson correlation coefficient \citep{Pearson1895}. However, we found in this study that this measure is too sensitive to extreme values and is therefore not robust enough for a pipeline implementation. Indeed, one outlier can result in a Pearson coefficient close to zero, even though all the other points are linearly correlated. We therefore propose the following modifications. We compute the standard deviation of the inversion coefficients and discard the coefficients that are not in the 3-sigma interval around zero. We chose to center our interval around zero because it worked well with our testing set by discarding the coefficients that we would have discarded manually. Alternatively, the interval can be centred around the mean of the coefficients, although with a smaller tolerance. We note that if the number of modes gets low (below 15) or if the inversion is based on the modes of one harmonic degree only, it is preferable to use the second option. Indeed, in such extreme conditions, the first criterion is unreliable and may discard a large fraction of the modes. We also note that up to two coefficients are typically discarded. In general, they correspond to the lowest radial order modes of the harmonic degrees. The correlation of the lag plot is then evaluated with the Spearman correlation coefficient \citep{Spearman1904}. This coefficient focuses on the rank variables $R(X)$ and $R(Y)$ instead of the random variables $X$ and $Y$ themselves. In that regard, the Spearman correlation coefficient is the Pearson correlation coefficient of the rank variables
\begin{align}
\mathcal{R} = \frac{\mathrm{cov}\left(R(X),R(Y)\right)}{\sigma_{R(X)}\cdot\sigma_{R(Y)}}.
\end{align}
This approach is more general, the Spearman coefficient indeed detects a monotonic correlation between the random variables, and has the great advantage of being significantly less sensitive to outliers. We note that if two random variables are linearly correlated, the Pearson and Spearman coefficients are equivalent. Because of these advantages, the Spearman coefficient is more robust and better suited for a pipeline implementation. As in JB23, we identified three regimes, which are summarised in Table \ref{tab:R_flag_regimes}. We consider that below $\mathcal{R}_t = 0.4$, the inversion coefficients show too much randomness for meaningful inversion. In that case, we reject the inversion result. If $\mathcal{R}_t > 0.65$, we consider that the inversion coefficients form smooth structures and we accept the inversion result. The in-between regime is more uncertain and we recommend to carry out further investigations. Because this test is based on the inversion coefficients, the boundaries of the different regimes are not dependent on the type of inversion that is considered.

\begin{table}
\centering
\caption{Instability regimes of the R-flag.}
\begin{tabular}{lcc}
\hline 
 & Criterion & R-flag \\ 
\hline \hline 
High instability & $\mathcal{R}_t < 0.4$  & reject \\ 
Moderate instability & $0.4 \leq\mathcal{R}_t\leq 0.65$ & check \\ 
Low instability & $\mathcal{R}_t > 0.65$ & accept \\ 
\hline 
\end{tabular} 
\label{tab:R_flag_regimes}
\end{table}

\begin{figure*}[htp!]
\centering
\begin{subfigure}[b]{.49\textwidth}
  \includegraphics[width=.99\textwidth]{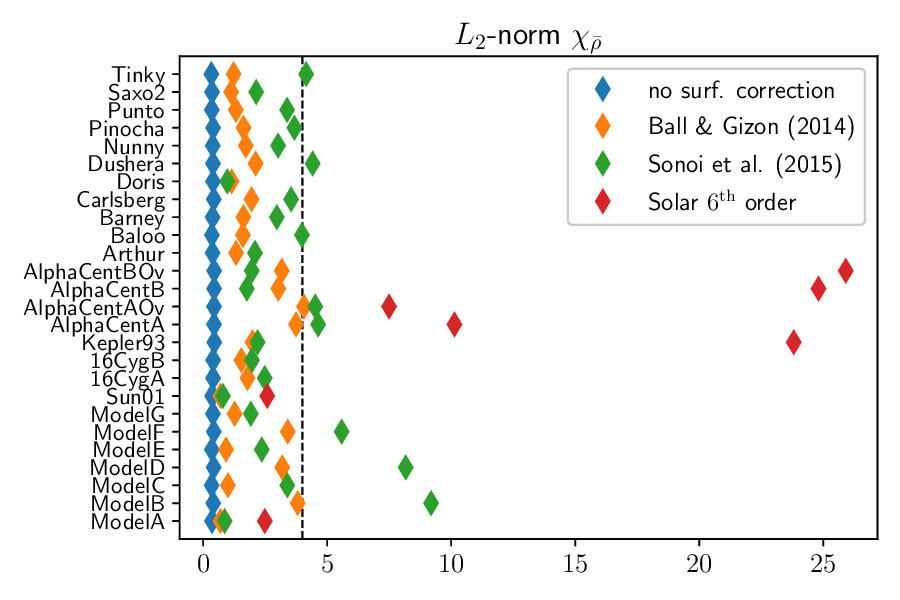}  
\end{subfigure}
\begin{subfigure}[b]{.49\textwidth}
  \includegraphics[width=.99\textwidth]{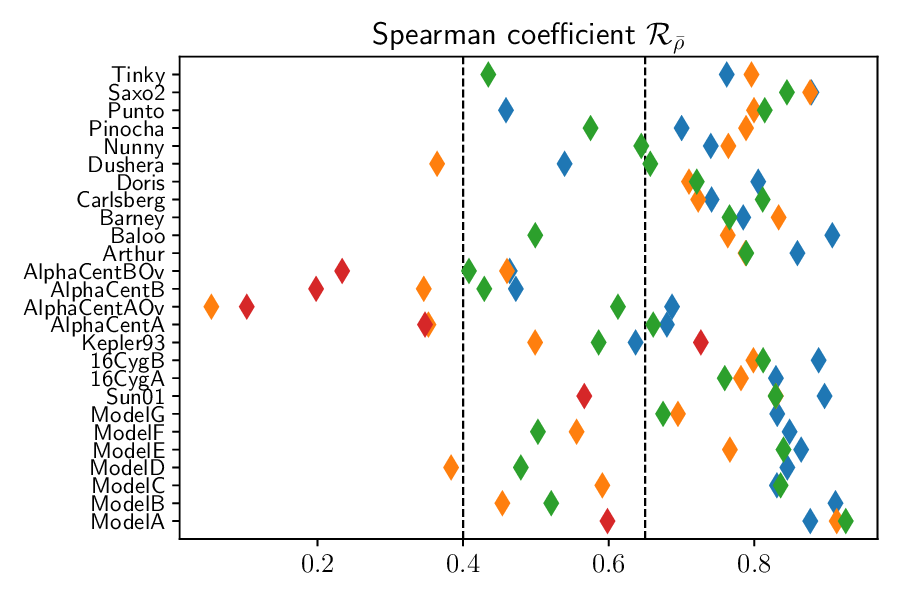}  
\end{subfigure}
\begin{subfigure}[b]{.49\textwidth}
  \includegraphics[width=.99\textwidth]{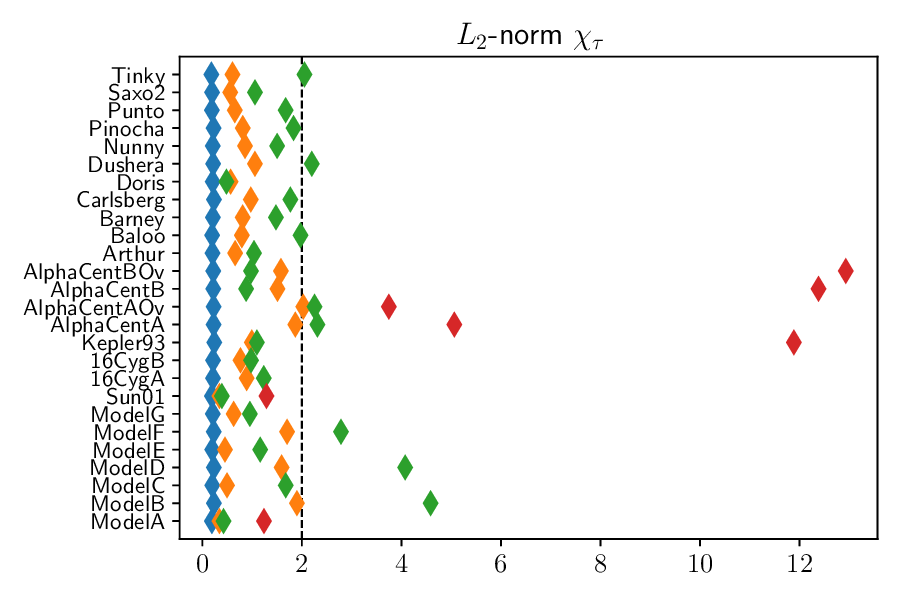}  
\end{subfigure}
\begin{subfigure}[b]{.49\textwidth}
  \includegraphics[width=.99\textwidth]{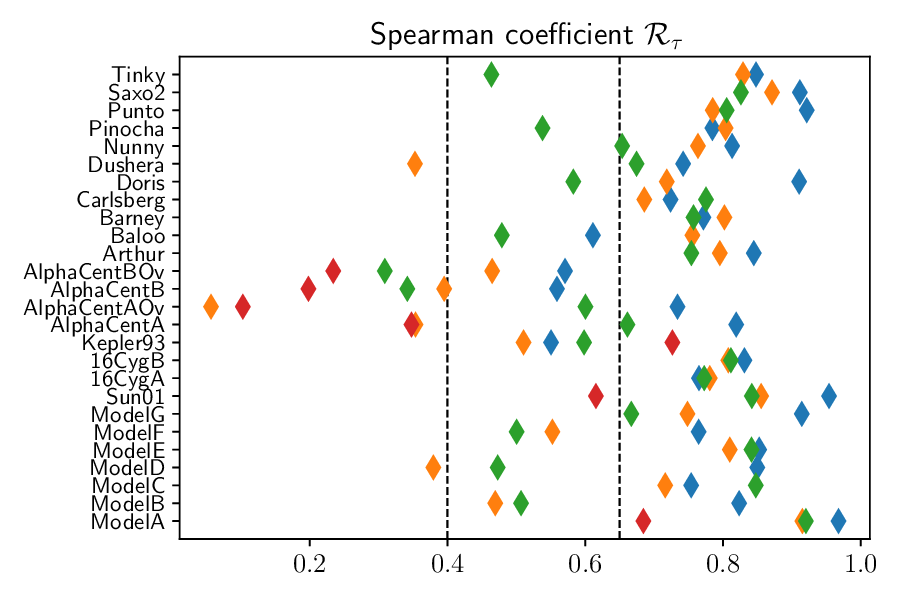}  
\end{subfigure}
\begin{subfigure}[b]{.49\textwidth}
  \includegraphics[width=.99\textwidth]{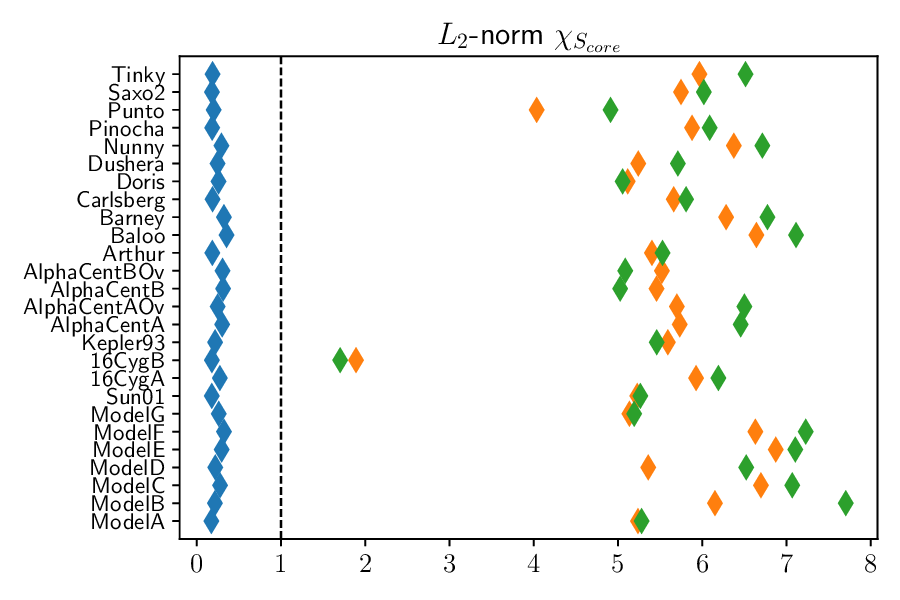}  
\end{subfigure}
\begin{subfigure}[b]{.49\textwidth}
  \includegraphics[width=.99\textwidth]{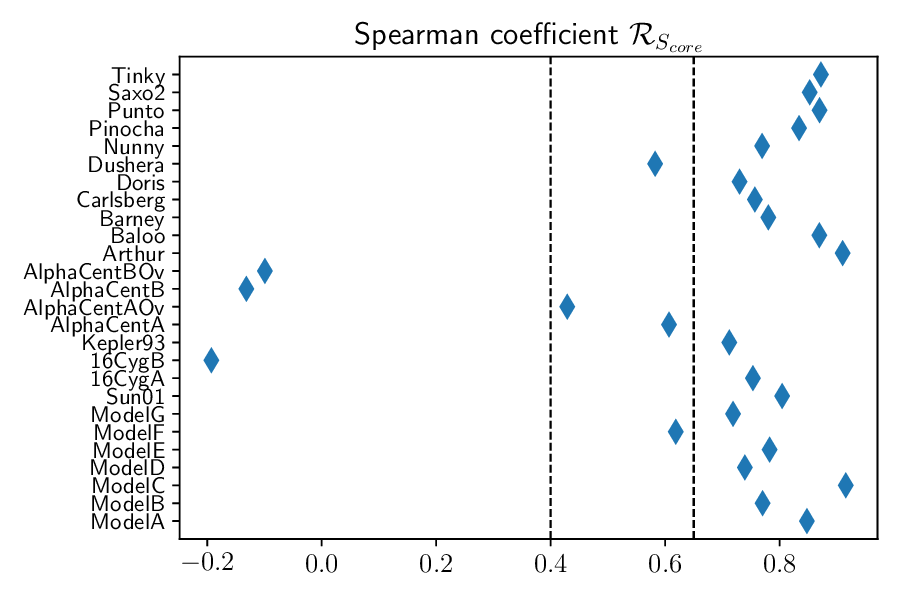}  
\end{subfigure}
\caption{Quality assessment results of the inversion carried out on our testing set by considering different surface effect prescriptions. \textit{Left column:} Quality of the fit of the target function by the averaging kernel quantified by $\chi_t$. \textit{Right column:} Spearman coefficient $\mathcal{R}_t$ of the lag plot. \textit{Top line:} Results of the mean density inversions. \textit{Middle line:} Results of the acoustic radius inversions. \textit{Bottom line:} Results of the central entropy inversions. The vertical dashed black lines delimits the different regimes of the selection flags.}
\label{fig:K&R_flags}
\end{figure*}

\section{Results}
\label{sec:results}

\subsection{Mean density and acoustic radius inversions}
\label{sec:results_rho_tau_inversions}
\begin{table*}[t!]
\centering
\caption{Results of our quality assessment procedure applied for the mean density inversions carried out on our testing set.}
\begin{tabular}{l|cc|cc|cc|cc}
\hline
 & \multicolumn{2}{c}{no surf.} & \multicolumn{2}{c}{BG14} & \multicolumn{2}{c}{TS15} & \multicolumn{2}{c}{solar $\rm6^{th}$ order} \\
Reference model & K-flag & R-flag & K-flag & R-flag & K-flag & R-flag & K-flag & R-flag \\
\hline \hline
ModelA & \cellcolor{lightgreen}accept & \cellcolor{lightgreen}accept & \cellcolor{lightgreen}accept & \cellcolor{lightgreen}accept & \cellcolor{lightgreen}accept & \cellcolor{lightgreen}accept & \cellcolor{lightgreen}accept & \cellcolor{cream}check \\ 
ModelB & \cellcolor{lightgreen}accept & \cellcolor{lightgreen}accept & \cellcolor{lightgreen}accept & \cellcolor{cream}check & \cellcolor{salmon}reject & \cellcolor{cream}check & - & - \\ 
ModelC & \cellcolor{lightgreen}accept & \cellcolor{lightgreen}accept & \cellcolor{lightgreen}accept & \cellcolor{cream}check & \cellcolor{lightgreen}accept & \cellcolor{lightgreen}accept & - & - \\ 
ModelD & \cellcolor{lightgreen}accept & \cellcolor{lightgreen}accept & \cellcolor{lightgreen}accept & \cellcolor{salmon}reject & \cellcolor{salmon}reject & \cellcolor{cream}check & - & - \\ 
ModelE & \cellcolor{lightgreen}accept & \cellcolor{lightgreen}accept & \cellcolor{lightgreen}accept & \cellcolor{lightgreen}accept & \cellcolor{lightgreen}accept & \cellcolor{lightgreen}accept & - & - \\ 
ModelF & \cellcolor{lightgreen}accept & \cellcolor{lightgreen}accept & \cellcolor{lightgreen}accept & \cellcolor{cream}check & \cellcolor{salmon}reject & \cellcolor{cream}check & - & - \\ 
ModelG & \cellcolor{lightgreen}accept & \cellcolor{lightgreen}accept & \cellcolor{lightgreen}accept & \cellcolor{lightgreen}accept & \cellcolor{lightgreen}accept & \cellcolor{lightgreen}accept & - & - \\ 
Sun01 &\cellcolor{lightgreen}accept & \cellcolor{lightgreen}accept & \cellcolor{lightgreen}accept & \cellcolor{lightgreen}accept & \cellcolor{lightgreen}accept & \cellcolor{lightgreen}accept & \cellcolor{lightgreen}accept & \cellcolor{cream}check \\ 
16CygA & \cellcolor{lightgreen}accept & \cellcolor{lightgreen}accept & \cellcolor{lightgreen}accept & \cellcolor{lightgreen}accept & \cellcolor{lightgreen}accept & \cellcolor{lightgreen}accept & - & - \\ 
16CygB & \cellcolor{lightgreen}accept & \cellcolor{lightgreen}accept & \cellcolor{lightgreen}accept & \cellcolor{lightgreen}accept & \cellcolor{lightgreen}accept & \cellcolor{lightgreen}accept & - & - \\ 
Kepler93 & \cellcolor{lightgreen}accept & \cellcolor{cream}check & \cellcolor{lightgreen}accept & \cellcolor{cream}check & \cellcolor{lightgreen}accept & \cellcolor{cream}check & \cellcolor{salmon}reject & \cellcolor{lightgreen}accept \\ 
AlphaCentA & \cellcolor{lightgreen}accept & \cellcolor{lightgreen}accept & \cellcolor{lightgreen}accept & \cellcolor{salmon}reject & \cellcolor{salmon}reject & \cellcolor{lightgreen}accept & \cellcolor{salmon}reject & \cellcolor{salmon}reject \\ 
AlphaCentAOv & \cellcolor{lightgreen}accept & \cellcolor{lightgreen}accept & \cellcolor{salmon}reject & \cellcolor{salmon}reject & \cellcolor{salmon}reject & \cellcolor{cream}check & \cellcolor{salmon}reject & \cellcolor{salmon}reject \\ 
AlphaCentB & \cellcolor{lightgreen}accept & \cellcolor{cream}check & \cellcolor{lightgreen}accept & \cellcolor{salmon}reject & \cellcolor{lightgreen}accept & \cellcolor{cream}check & \cellcolor{salmon}reject & \cellcolor{salmon}reject \\ 
AlphaCentBOv & \cellcolor{lightgreen}accept & \cellcolor{cream}check & \cellcolor{lightgreen}accept & \cellcolor{cream}check & \cellcolor{lightgreen}accept & \cellcolor{cream}check & \cellcolor{salmon}reject & \cellcolor{salmon}reject \\ 
Arthur & \cellcolor{lightgreen}accept & \cellcolor{lightgreen}accept & \cellcolor{lightgreen}accept & \cellcolor{lightgreen}accept & \cellcolor{lightgreen}accept & \cellcolor{lightgreen}accept & - & - \\ 
Baloo & \cellcolor{lightgreen}accept & \cellcolor{lightgreen}accept & \cellcolor{lightgreen}accept & \cellcolor{lightgreen}accept & \cellcolor{lightgreen}accept & \cellcolor{cream}check & - & - \\ 
Barney & \cellcolor{lightgreen}accept & \cellcolor{lightgreen}accept & \cellcolor{lightgreen}accept & \cellcolor{lightgreen}accept & \cellcolor{lightgreen}accept & \cellcolor{lightgreen}accept & - & - \\ 
Carlsberg & \cellcolor{lightgreen}accept & \cellcolor{lightgreen}accept & \cellcolor{lightgreen}accept & \cellcolor{lightgreen}accept & \cellcolor{lightgreen}accept & \cellcolor{lightgreen}accept & - & - \\ 
Doris & \cellcolor{lightgreen}accept & \cellcolor{lightgreen}accept & \cellcolor{lightgreen}accept & \cellcolor{lightgreen}accept & \cellcolor{lightgreen}accept & \cellcolor{lightgreen}accept & - & - \\ 
Dushera & \cellcolor{lightgreen}accept & \cellcolor{cream}check & \cellcolor{lightgreen}accept & \cellcolor{salmon}reject & \cellcolor{salmon}reject & \cellcolor{lightgreen}accept & - & - \\ 
Nunny & \cellcolor{lightgreen}accept & \cellcolor{lightgreen}accept & \cellcolor{lightgreen}accept & \cellcolor{lightgreen}accept & \cellcolor{lightgreen}accept & \cellcolor{cream}check & - & - \\ 
Pinocha & \cellcolor{lightgreen}accept & \cellcolor{lightgreen}accept & \cellcolor{lightgreen}accept & \cellcolor{lightgreen}accept & \cellcolor{lightgreen}accept & \cellcolor{cream}check & - & - \\ 
Punto & \cellcolor{lightgreen}accept & \cellcolor{cream}check & \cellcolor{lightgreen}accept & \cellcolor{lightgreen}accept & \cellcolor{lightgreen}accept & \cellcolor{lightgreen}accept & - & - \\ 
Saxo2 & \cellcolor{lightgreen}accept & \cellcolor{lightgreen}accept & \cellcolor{lightgreen}accept & \cellcolor{lightgreen}accept & \cellcolor{lightgreen}accept & \cellcolor{lightgreen}accept & - & - \\ 
Tinky & \cellcolor{lightgreen}accept & \cellcolor{lightgreen}accept & \cellcolor{lightgreen}accept & \cellcolor{lightgreen}accept & \cellcolor{salmon}reject & \cellcolor{cream}check & - & - \\ 
\hline 
\end{tabular}
{\par\small\justify\textbf{Notes.}  We considered four semi-empirical surface effect prescriptions: no surface corrections (no surf.), \citet{Ball&Gizon2014} two terms (BG14), \citet{Sonoi2015} two terms (TS15), and a sixth order polynomial (solar $6^{\mathrm{th}}$ order). \par}
\label{tab:rho_flags}
\end{table*}

The mean density inversion and the acoustic radius inversion are based on the same structural kernels and share the same trade-off parameters. The form of their seismic indicator is simple compared to more ambitious inversions such as the central entropy inversion. Due to these similarities, the results of our quality assessment procedure are very similar. In that regard, if the mean density inversion is flagged as accepted, the corresponding acoustic radius inversion is typically flagged as accepted too, and vice versa. In this section, we therefore focus our discussion on the results of the mean density inversions, which are displayed in the top line of Fig. \ref{fig:K&R_flags} and in Table \ref{tab:rho_flags}, and the results of the acoustic radius inversions can be found in Appendix \ref{appendix:supp_data_acoustic_radius_inversion}.

The results of the calibrator targets are consistent with our expectations. For the solar model, introducing additional free parameters to describe the surface effects increases the value of the $\chi_{\bar{\rho}}$ quantifier. The opposite behaviour is observed with the second quantifier, the Spearman correlation coefficient $\mathcal{R}_{\bar{\rho}}$. In addition, all these inversions correct towards the expected mean density range. The results of 16 Cyg A and B behave similarly to the results of the Sun. The inversions of both binary components correct towards the measurements of \citet{Buldgen2022b}, and all these inversions are accurately flagged as accepted. In addition, with the high data quality of these targets, both quality quantifiers correctly reflect that the \citet{Ball&Gizon2014} prescription is slightly more stable than the \citet{Sonoi2015} prescription, and that the inversions that neglect surface effects are the most stable. As a sidenote, we note that neglecting surface effects gives the most stable inversions because it imposes fewer free variables in the minimisations, but it does not mean that the outcome of this inversion is the best physical results. Indeed, as pointed out by JB23 and by many other studies, the \citet{Ball&Gizon2014} prescription is the best default choice. For the $\alpha$ Cen binary system, we expected poor quality inversion results. Indeed, the data quality of these targets is lower than the data quality of the other calibrator targets. For these targets, we investigated two types of reference models, with or without overshooting in $\alpha$ Cen A. From the literature, we know that the relative frequency differences of $\alpha$ Cen B are too large for a robust inversion based on individual frequencies, and that the inversion results of the model $\alpha$ Cen A including overshooting are significantly affected by the choice of the mode set, suggesting that some of the modes have a non-linear character. As expected, the inversions using the sixth order polynomial to describe the surface effects are rejected. The rest of the inversions are either flagged as rejected or as requiring a manual and thorough investigation. The results for these targets show the relevance of using both quality flags. Indeed, due to the amplitude differences of the target function of models over a large mass range, the fine-tuning of the rejection threshold of the K-flag is limited, and for these models in particular, it would benefit from a lower tolerance. Although the K-flag has a non-negligible false positive rate, most of these problematic inversions are detected by the second quality flag. In addition, we note that for the model of $\alpha$ Cen A without overshooting and with the \citet{Sonoi2015} surface effects prescription, the inversion result is rejected by the K-flag but not by the R-flag, which illustrates a limitation of the R-flag. If the target function is not reproduced at all by the averaging kernel, the correction proposed by the inversion is non-physical and is the result of the poor fit of the target function. However, it may still create structures in the inversion coefficients that are detected by the R-flag. This is not an issue for our assessment procedure, where the K-flag is first computed. Indeed, if the inversion is rejected by the K-flag, it is unnecessary to compute the R-flag. The quality assessment results of Kepler-93 also correspond to our expectations. The data quality of this target is lower than the data quality of the best \emph{Kepler} targets with the \citet{Ball&Gizon2014} and \citet{Sonoi2015} prescriptions, and the target function is therefore less well reproduced by the averaging kernel. This is detected by the R-flag that labelled these inversions as requiring a manual check. The data quality of Kepler-93 is insufficient for a surface effect prescriptions with six free variables. This inversion is rejected by the K-flag, but not by the R-flag, which again shows that the R-flag should not be computed for inversion results that were rejected by the K-flag. 

The results of the second category of targets are shown in Fig. \ref{fig:K&R_flags} and in Table \ref{tab:rho_flags}. The most problematic cases are directly discarded by the K-flag, and the R-flag point out the lower quality inversions. Indeed, by checking manually the fit of the averaging kernel, we identified model B, C, D, and F, and Dushera as lower quality inversion, which are also correctly highlighted by our assessment procedure. The result for Dushera is particularly interesting. Despite having lower data quality than the best targets (e.g. Doris), the data quality is similar to that of other stable inversion results (e.g. Pinocha). However, the inversion result of Dushera is flagged as unreliable. As in the case of $\alpha$ Cen A, it is possible that one (or more) of the modes is affected by non-linearities, which could explain this unstable behaviour. Alternatively, it is also possible that there was an issue with the peak bagging. Indeed, \citet{Roxburgh2017} observed anomalies in some of the \emph{LEGACY} data. Although this study did not analyse Dushera's data, it is possible that it was impacted by the same issues, which could also explain the unstable behaviour of the inversion. Both possibilities could be investigated by a comprehensive analysis using local minimisations, which is beyond the scope of this study. Nevertheless, this result is promising as it indicates that our assessment procedure is able to highlight problematic inversion results that are usually difficult to detect manually. Additionally, the stable inversion results are correctly flagged as stable. Hence, the combination of both flags performs satisfactorily for the mean density inversion and the acoustic radius inversion with respect to a human modeller.

The \citet{Ball&Gizon2014} prescription is the preferred surface effect prescription for the PLATO pipeline and it is therefore relevant to look at the flag distribution of the inversion results with this prescription. We note that our testing set is not bias-free, we only considered medium to high data-quality targets (more than 30 observed modes) and we included several poor quality inversion results to verify that they could be spotted by our assessment procedure. The percentages that we quote below should therefore be interpreted with caution, and further investigations with a larger statistics and including lower data-quality targets are required. With our testing set, about 20\% of the results are flagged as rejected, about 20\% as to be checked manually, and the remaining 60\% as accepted. Although it is difficult to draw robust conclusions based on these numbers, we note that few inversion results are rejected and also that few results require further investigations. This is an important aspect because such investigations cannot be carried out within the pipeline. Hence, these results comfort us in the idea that the mean density inversion is suited for a large-scale application.

\subsection{Central entropy inversion}
\label{sec:results_score_inversion}
The results of the central entropy inversion are shown in the bottom line of Fig. \ref{fig:K&R_flags} and in Table \ref{tab:Score_flags}. For all the models, we found that the inversion fails if surface effects are included. Indeed, the inversions using the \citet{Ball&Gizon2014} and \citet{Sonoi2015} prescriptions have averaging kernels that completely miss the central stellar features of the target function. Hence, all these inversion results can be discarded because these central layers are the region of interest of the inversion. In addition, the situation is even worse with the sixth order polynomial. In this configuration, the number of degrees of freedom is insufficient to carry out the SOLA inversion. As expected, the K-flag rejects all these inversions. Although we recommend to avoid computing the R-flag of inversions that were rejected by the K-flag, we provided in Table \ref{tab:Score_flags} the R-flag of such models to illustrate why we emitted this recommendation. As shown in Table~\ref{tab:Score_flags}, the R-flag is not reliable in such conditions. Regarding the results of the inversions that neglect surface effects, our quality assessment procedure performs equivalently to a human modeller. The models with a lower inversion quality are indeed correctly spotted by the R-flag. 

These results however question the relevance of including this inversion in a pipeline, at least in its current form. This indicator is indeed designed to probe the central stellar layers, but it is at the same time very sensitive to the surface regions because it is based on the $S_{5/3}$ profile which is too sensitive to these regions. Hence, to robustly interpret the results of this type of inversion, it is necessary to generate a set of models representative of the observed target and study the behaviour of the inversion on the set, as is done in \citet{Buldgen2017_proceeding}, \citet{Salmon2021} and \citet{Buldgen2022b} for example. We note that using a similar indicator but based on frequency separation ratios \citep{Betrisey&Buldgen2022} would also be incompatible with a pipeline approach. Even though such an indicator is significantly less affected by surface effects, it is based on ratios which might take very small values, and therefore result in singular relative ratios differences. This inversion therefore requires some caution in the data processing and in the interpretation of the results. For this inversion too, it is necessary to generate a set of models and study the inversion on the set. 

\begin{table*}[t!]
\centering
\caption{Results of our quality assessment procedure applied for the central entropy inversions carried out on our testing set.}
\begin{tabular}{l|cc|cc|cc}
\hline
 & \multicolumn{2}{c}{no surf.} & \multicolumn{2}{c}{BG14} & \multicolumn{2}{c}{TS15} \\
Reference model & K-flag & R-flag & K-flag & R-flag & K-flag & R-flag \\
\hline \hline
ModelA & \cellcolor{lightgreen}accept & \cellcolor{lightgreen}accept & \cellcolor{salmon}reject & \cellcolor{lightgreen}accept & \cellcolor{salmon}reject & \cellcolor{lightgreen}accept \\ 
ModelB & \cellcolor{lightgreen}accept & \cellcolor{lightgreen}accept & \cellcolor{salmon}reject & \cellcolor{lightgreen}accept & \cellcolor{salmon}reject & \cellcolor{lightgreen}accept \\ 
ModelC & \cellcolor{lightgreen}accept & \cellcolor{lightgreen}accept & \cellcolor{salmon}reject & \cellcolor{lightgreen}accept & \cellcolor{salmon}reject & \cellcolor{lightgreen}accept \\ 
ModelD & \cellcolor{lightgreen}accept & \cellcolor{lightgreen}accept & \cellcolor{salmon}reject & \cellcolor{cream}check & \cellcolor{salmon}reject & \cellcolor{cream}check \\ 
ModelE & \cellcolor{lightgreen}accept & \cellcolor{lightgreen}accept & \cellcolor{salmon}reject & \cellcolor{cream}check & \cellcolor{salmon}reject & \cellcolor{cream}check \\ 
ModelF & \cellcolor{lightgreen}accept & \cellcolor{cream}check & \cellcolor{salmon}reject & \cellcolor{cream}check & \cellcolor{salmon}reject & \cellcolor{cream}check \\ 
ModelG & \cellcolor{lightgreen}accept & \cellcolor{lightgreen}accept & \cellcolor{salmon}reject & \cellcolor{lightgreen}accept & \cellcolor{salmon}reject & \cellcolor{lightgreen}accept \\ 
Sun01 & \cellcolor{lightgreen}accept & \cellcolor{lightgreen}accept & \cellcolor{salmon}reject & \cellcolor{lightgreen}accept & \cellcolor{salmon}reject & \cellcolor{lightgreen}accept \\ 
16CygA & \cellcolor{lightgreen}accept & \cellcolor{lightgreen}accept & \cellcolor{salmon}reject & \cellcolor{lightgreen}accept & \cellcolor{salmon}reject & \cellcolor{lightgreen}accept \\ 
16CygB & \cellcolor{lightgreen}accept & \cellcolor{salmon}reject & \cellcolor{salmon}reject & \cellcolor{salmon}reject & \cellcolor{salmon}reject & \cellcolor{salmon}reject \\ 
Kepler93 & \cellcolor{lightgreen}accept & \cellcolor{lightgreen}accept & \cellcolor{salmon}reject & \cellcolor{lightgreen}accept & \cellcolor{salmon}reject & \cellcolor{cream}check \\ 
AlphaCentA & \cellcolor{lightgreen}accept & \cellcolor{cream}check & \cellcolor{salmon}reject & \cellcolor{cream}check & \cellcolor{salmon}reject & \cellcolor{salmon}reject \\ 
AlphaCentAOv & \cellcolor{lightgreen}accept & \cellcolor{cream}check & \cellcolor{salmon}reject & \cellcolor{salmon}reject & \cellcolor{salmon}reject & \cellcolor{salmon}reject \\ 
AlphaCentB & \cellcolor{lightgreen}accept & \cellcolor{salmon}reject & \cellcolor{salmon}reject & \cellcolor{salmon}reject & \cellcolor{salmon}reject & \cellcolor{salmon}reject \\ 
AlphaCentBOv & \cellcolor{lightgreen}accept & \cellcolor{salmon}reject & \cellcolor{salmon}reject & \cellcolor{salmon}reject & \cellcolor{salmon}reject & \cellcolor{salmon}reject \\ 
Arthur & \cellcolor{lightgreen}accept & \cellcolor{lightgreen}accept & \cellcolor{salmon}reject & \cellcolor{lightgreen}accept & \cellcolor{salmon}reject & \cellcolor{lightgreen}accept \\ 
Baloo & \cellcolor{lightgreen}accept & \cellcolor{lightgreen}accept & \cellcolor{salmon}reject & \cellcolor{cream}check & \cellcolor{salmon}reject & \cellcolor{lightgreen}accept \\ 
Barney & \cellcolor{lightgreen}accept & \cellcolor{lightgreen}accept & \cellcolor{salmon}reject & \cellcolor{lightgreen}accept & \cellcolor{salmon}reject & \cellcolor{lightgreen}accept \\ 
Carlsberg & \cellcolor{lightgreen}accept & \cellcolor{lightgreen}accept & \cellcolor{salmon}reject & \cellcolor{lightgreen}accept & \cellcolor{salmon}reject & \cellcolor{lightgreen}accept \\ 
Doris & \cellcolor{lightgreen}accept & \cellcolor{lightgreen}accept & \cellcolor{salmon}reject & \cellcolor{lightgreen}accept & \cellcolor{salmon}reject & \cellcolor{cream}check \\ 
Dushera & \cellcolor{lightgreen}accept & \cellcolor{cream}check & \cellcolor{salmon}reject & \cellcolor{cream}check & \cellcolor{salmon}reject & \cellcolor{cream}check \\ 
Nunny & \cellcolor{lightgreen}accept & \cellcolor{lightgreen}accept & \cellcolor{salmon}reject & \cellcolor{lightgreen}accept & \cellcolor{salmon}reject & \cellcolor{lightgreen}accept \\ 
Pinocha & \cellcolor{lightgreen}accept & \cellcolor{lightgreen}accept & \cellcolor{salmon}reject & \cellcolor{lightgreen}accept & \cellcolor{salmon}reject & \cellcolor{salmon}reject \\ 
Punto & \cellcolor{lightgreen}accept & \cellcolor{lightgreen}accept & \cellcolor{salmon}reject & \cellcolor{lightgreen}accept & \cellcolor{salmon}reject & \cellcolor{lightgreen}accept \\ 
Saxo2 & \cellcolor{lightgreen}accept & \cellcolor{lightgreen}accept & \cellcolor{salmon}reject & \cellcolor{lightgreen}accept & \cellcolor{salmon}reject & \cellcolor{cream}check \\ 
Tinky & \cellcolor{lightgreen}accept & \cellcolor{lightgreen}accept & \cellcolor{salmon}reject & \cellcolor{lightgreen}accept & \cellcolor{salmon}reject & \cellcolor{salmon}reject \\ 
\hline 
\end{tabular}
{\par\small\justify\textbf{Notes.}  We considered three semi-empirical surface effect prescriptions: no surface corrections (no surf.), \citet{Ball&Gizon2014} two terms (BG14), and \citet{Sonoi2015} two terms (TS15). \par}
\label{tab:Score_flags}
\end{table*}

\section{Discussion}
\label{sec:discussion}

\subsection{Preconditioning of the inversion}
\label{sec:preconditioning_inversion}
The variational inversions are based on a linear formalism. During the derivation of the structure inversion equation at the basis of the variational inversions, this linearity assumption allows us to neglect of lot of higher order terms, notably surface terms arising from partial integration, and end up with a simple equation directly relating frequency differences to structure differences. Non-linearities may therefore significantly affect the inversion by inducing unwanted compensations, and they are usually difficult to spot. In this section, we discuss two types of common non-linearities, the mode non-linearity and the non-linear regime of the reference model.

In the first scenario, the mode itself exhibits a non-linear behaviour. A mixed mode mistaken for a pressure mode fits within this category. It is often tricky to detect and in our testing set, we suspect that the model of $\alpha$ Cen A including overshooting is affected by such a non-linearity. For the rest of the targets, there is a priori no sign of mode non-linearity. We note that it would require a thorough investigation of each target to robustly disprove the presence of such non-linearities, but based on the current inversion results, it is reasonable to assume that it only affects a minority of targets. It is therefore unlikely to be an issue in a pipeline.

\begin{figure}[t!]
\centering
\includegraphics[scale=0.5]{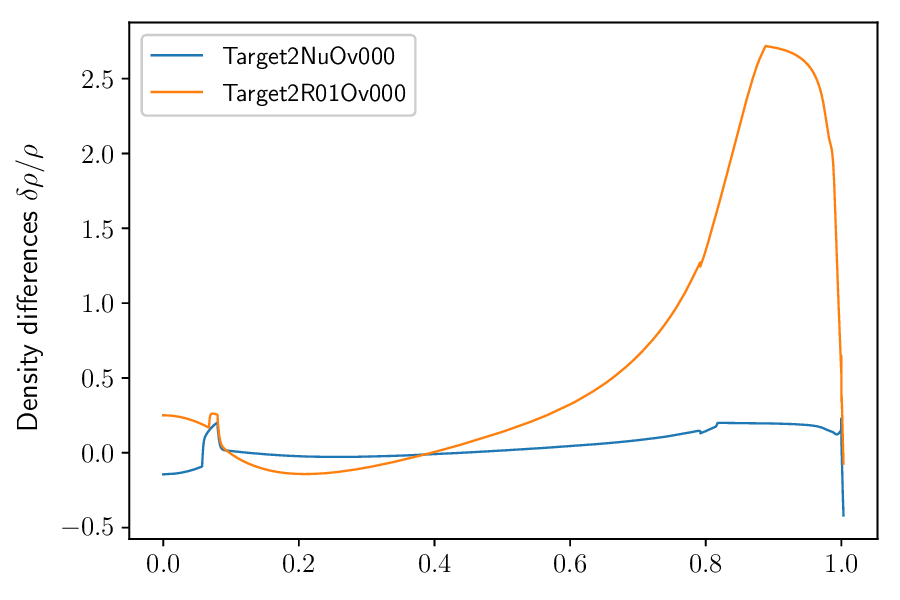} 
\caption{Density differences of two reference models of target 2 from \citet{Betrisey&Buldgen2022}. The reference model `target2NuOv000' is within the linear regime, while the model `target2R01Ov000' is outside of the linear regime.}
\label{fig:density_differences}
\end{figure}

The second scenario is related to the reference model. If it is too far away from the observed target in the parameter space, structural differences may be too large for the linear assumption and might induce compensations in the inversion. In Fig. \ref{fig:density_differences}, we show the structural density differences for a model within the linear regime (`target2NuOv000'), fitting the individual frequencies, and for a model outside the linear regime (`target2R01Ov000'), fitting the $r_{01}$ ratios alone.  We took these models from \citet{Betrisey&Buldgen2022}. In the illustration, `target2R01Ov000' has large differences in the upper layers which are magnified by the structural kernels and their large amplitude in these regions. It induces unwanted compensations and the inversion is unsuccessful. The boundaries of the linear regime are often unclear and may change from target to target. However, good preconditioning can ensure that the reference model is in the linear regime. Hence, a fit of the individual frequencies and the classical constraints with a MCMC typically ensures this assumption. This type of non-linearity is therefore not an issue for the modelling strategy that was proposed in JB23, which starts with such a fit and then corrects for the surface effects by combining a mean density inversion and a fit of frequency separation ratios.

\subsection{Limited mode sets}
\label{sec:limited_mode_sets}
\begin{table*}[t!]
\centering
\caption{Assessment flags of the tests on limited mode sets. We carried out mean density inversions and considered four semi-empirical surface effect prescriptions: no surface corrections (no surf.), \citet{Ball&Gizon2014} (BG14), \citet{Sonoi2015} (TS15), and a sixth order polynomial (solar $6^{\mathrm{th}}$ order).}
\resizebox{\linewidth}{!}{
\begin{tabular}{lcc|cc|cc|cc|cc}
\hline
 & & & \multicolumn{2}{c}{no surf.} & \multicolumn{2}{c}{BG14} & \multicolumn{2}{c}{TS15} & \multicolumn{2}{c}{solar $\rm6^{th}$ order} \\
Reference model & Mode number & Mode set & K-flag & R-flag & K-flag & R-flag & K-flag & R-flag & K-flag & R-flag \\
\hline \hline
Sun01 & 54 & $l=0-2$ &\cellcolor{lightgreen}accept & \cellcolor{lightgreen}accept & \cellcolor{lightgreen}accept & \cellcolor{lightgreen}accept & \cellcolor{lightgreen}accept & \cellcolor{lightgreen}accept & \cellcolor{lightgreen}accept & \cellcolor{cream}check \\ 
Sun01 & 18 & $l=0$ & \cellcolor{lightgreen}accept & \cellcolor{lightgreen}accept & \cellcolor{lightgreen}accept & \cellcolor{lightgreen}accept & \cellcolor{lightgreen}accept & \cellcolor{lightgreen}accept & \cellcolor{lightgreen}accept & \cellcolor{salmon}reject \\ 
Sun01 & 10 & $l=0, n=17-26$ & \cellcolor{lightgreen}accept & \cellcolor{cream}check & \cellcolor{salmon}reject & - & \cellcolor{salmon}reject & - & \cellcolor{salmon}reject & - \\ 
Sun01 & 12 & $l=0-2,n=19-22$ & \cellcolor{lightgreen}accept & \cellcolor{salmon}reject & \cellcolor{salmon}reject & - & \cellcolor{salmon}reject & - & \cellcolor{salmon}reject & - \\
Kepler93 & 32 & $l=0-2$ & \cellcolor{lightgreen}accept & \cellcolor{cream}check & \cellcolor{lightgreen}accept & \cellcolor{cream}check & \cellcolor{lightgreen}accept & \cellcolor{cream}check & \cellcolor{salmon}reject & - \\ 
Kepler93 & 12 & $l=0$ & \cellcolor{lightgreen}accept & \cellcolor{cream}check & \cellcolor{lightgreen}accept & \cellcolor{lightgreen}accept & \cellcolor{lightgreen}accept & \cellcolor{lightgreen}accept & \cellcolor{salmon}reject & - \\
AlphaCentA & 44 & $l=0-3$ & \cellcolor{lightgreen}accept & \cellcolor{lightgreen}accept & \cellcolor{lightgreen}accept & \cellcolor{salmon}reject & \cellcolor{salmon}reject & \cellcolor{lightgreen}accept & \cellcolor{salmon}reject & - \\ 
AlphaCentA & 10 & $l=0$ & \cellcolor{lightgreen}accept & \cellcolor{cream}check & \cellcolor{salmon}reject & - & \cellcolor{salmon}reject & - & \cellcolor{salmon}reject & - \\
\hline 
\end{tabular}
}
\label{tab:rho_flags_limited_dataset}
\end{table*}

In Sec. \ref{sec:quality_assessment_procedure}, we tested our quality assessment procedure on targets with a data quality going from medium to high. It corresponds to mode sets that are composed of more than 30 individual modes. This number of modes allowed us to use a statistical tool at the basis of the R-flag. PLATO will however detect many targets with fewer pulsation frequencies. Hence, we tested how our assessment procedure behaves in such cases. We investigated three calibrator targets, the Sun, Kepler-93, and $\alpha$ Cen A, that are representative of good, medium, and poor quality inversions, respectively. 

We summarised the results of our assessment procedure in Table~\ref{tab:rho_flags_limited_dataset}. Due to the lower number of modes, we did not discard inversion coefficient above the $3\sigma$ threshold used to remove the outliers before the computation of the Spearman correlation coefficient. For the Sun, the inversion is robust if only the $l=0$ modes are used (18 modes in total), except for the sixth order surface effect prescription, as expected. We also tested carrying out the inversion based on ten $l=0$ modes around $\nu_{\mathrm{max}}$ only and on four modes around $\nu_{\mathrm{max}}$ of each harmonic degree (12 modes in total). In such conditions, the fit of the target function by the averaging kernel is insufficient and the inversion is rejected by our assessment procedure. For Kepler-93, the quality of the fit of the target function by the averaging kernel with the \citet{Ball&Gizon2014} and \citet{Sonoi2015} prescriptions is in a grey zone. Just based on this fit, we would have rejected the inversion results. However, based on the inverted mean densities that are consistent with \citet{Betrisey2022}, and based on the inversion coefficients that form smooth structures, it seems that these inversions were successful. As expected, these inversions are flagged as accepted by our assessment procedure. Unsurprisingly, $\alpha$ Cen A, which had poor quality inversion results with all the modes, has even worse quality inversion results if only the $l=0$ modes are used.

Theoretically, an inversion based on a dozen modes is not expected to be challenging. It assumes that theses modes were carefully selected nonetheless. This is doable in an hare and hounds exercise, but as we have seen in this section, the outcomes of such inversions are unpredictable. Indeed an actual mode set is composed of the modes that were detected by the instrument, and there is no possibility to select carefully the modes. The inversion carried out on the $l=0$ modes of Kepler-93 was successful but there is no guarantee that it will be the case for another target.


\section{Conclusions}
\label{sec:conclusions}
In Sec. \ref{sec:modelling_strategy}, we presented the inversion types that we considered in this study and we also presented our testing set. In Sec. \ref{sec:quality_assessment_procedure}, we described our quality assessment procedure which was then applied on our testing set in Sec. \ref{sec:results}. Finally in Sec. \ref{sec:discussion}, we discussed best practices to consider for a large scale application of our assessment procedure.

Even though the mean density inversion \citep{Reese2012} and the acoustic radius inversion \citep{Buldgen2015a} were originally developed for individual modelling, the results of JB23 and of this study comfort us in the idea that these inversions are compatible with a large-scale application. The central entropy inversion \citep{Buldgen2018}, which is based on a seismic indicator with a more complex form, is however not compatible with a large-scale application in its current form. We found that our procedure performs equally well as a human modeller. Nonetheless, we note that we mainly tested our procedure on medium to high quality targets, with at least 30 observed modes. Dealing with lower statistics may be an issue for the second test of our procedure, but not for the first test. In that regard, we believe that our procedure is still applicable on limited mode sets, although this aspect would benefit from further investigations. However, a limited mode set of a dozen of frequencies could be an issue for the inversion itself. Indeed, the kernels of such mode sets may be insufficient for the averaging kernel to reproduce the target function. In these conditions, the success of an inversion becomes unpredictable and is sensitive to the mode set which is used.

Putting these results in the context of the PLATO mission, our quality assessment procedure of seismic inversions showed promising results. It is indeed based on by-products of the inversion and the two quality tests which are performed require few numerical resources. Hence, our assessment procedure can assess quickly and inexpensively the quality of an inversion, while still performing as well as a human modeller.


\section*{Acknowledgements}
J.B. and G.B. acknowledge funding from the SNF AMBIZIONE grant No 185805 (Seismic inversions and modelling of transport processes in stars). G.M. has received funding from the European Research Council (ERC) under the European Union's Horizon 2020 research and innovation programme (grant agreement No 833925, project STAREX). Finally, this work has benefited from financial support by CNES (Centre National des Etudes Spatiales) in the framework of its contribution to the PLATO mission.


\bibliography{bibliography.bib}

\begin{appendix}

\section{Supplementary data for K-flag}
\begin{figure}[h!]
\centering
\begin{subfigure}[b]{.43\textwidth}
  \includegraphics[width=.99\textwidth]{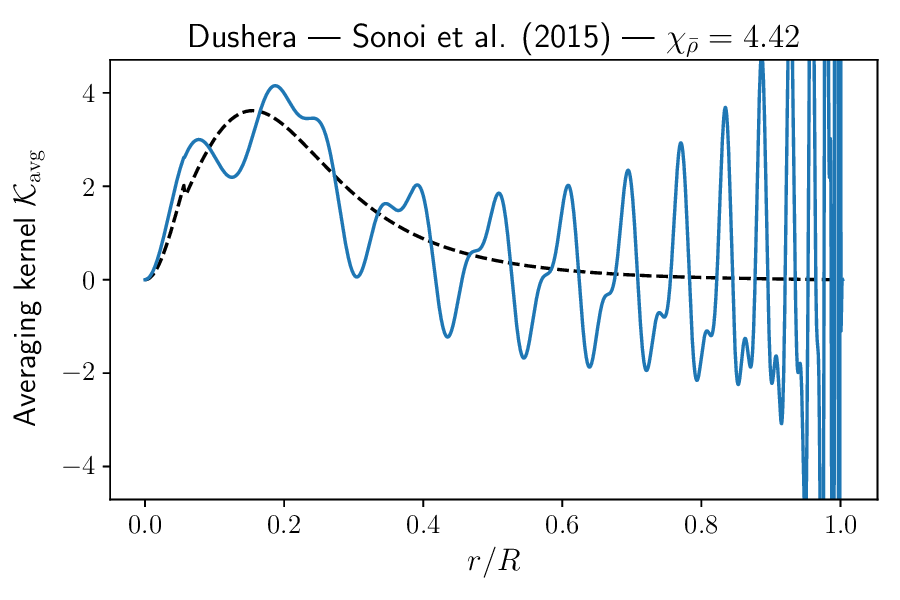}  
\end{subfigure}
\begin{subfigure}[b]{.43\textwidth}
  \includegraphics[width=.99\textwidth]{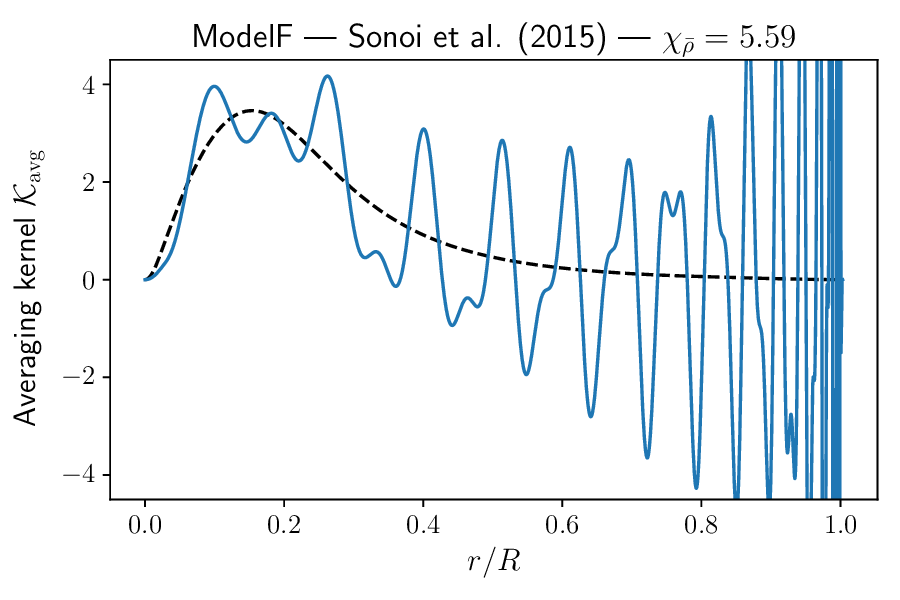}  
\end{subfigure} 
\begin{subfigure}[b]{.43\textwidth}
  \includegraphics[width=.99\textwidth]{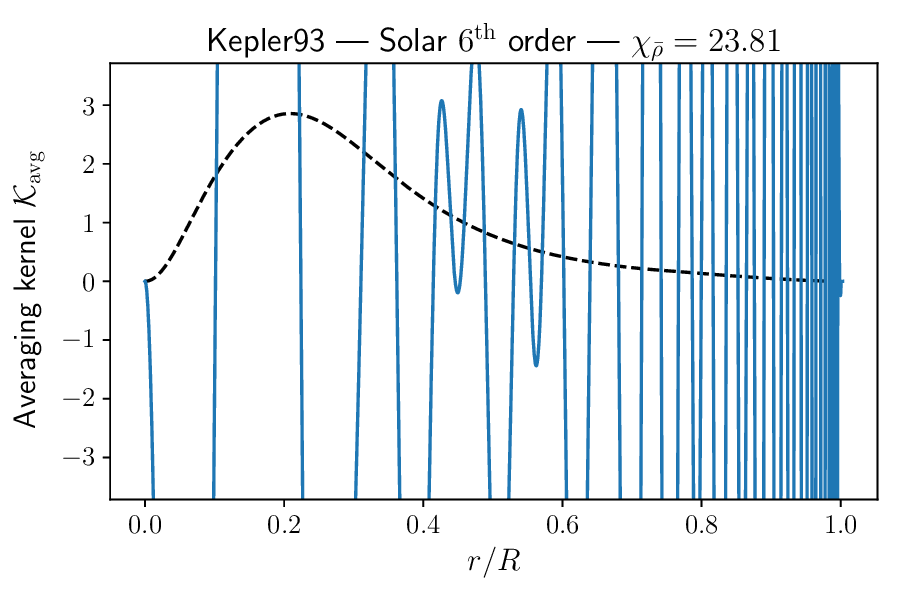}  
\end{subfigure} 
\caption{Outlying cases where the averaging kernel is unable to reproduce the target function. The $\chi_{\bar{\rho}}$ quantifier therefore takes a large value. The averaging kernel is shown in blue and the target function is the dashed black line.}
\label{fig:Kflag_abberant_cases}

\end{figure}

In Fig. \ref{fig:Kflag_abberant_cases}, we show illustrations of outliers which are directly rejected by the first flag of our quality assessment procedure.

\label{appendix:supp_data_Kflag}

\section{Assessment flags of the acoustic radius inversions}
\label{appendix:supp_data_acoustic_radius_inversion}
\begin{table*}[t!]
\centering
\caption{Results of our quality assessment procedure applied for the acoustic radius inversions carried out on our testing set.}
\begin{tabular}{l|cc|cc|cc|cc}
\hline
 & \multicolumn{2}{c}{no surf.} & \multicolumn{2}{c}{BG14} & \multicolumn{2}{c}{TS15} & \multicolumn{2}{c}{solar $\rm6^{th}$ order} \\
Reference model & K-flag & R-flag & K-flag & R-flag & K-flag & R-flag & K-flag & R-flag \\
\hline \hline
ModelA & \cellcolor{lightgreen}accept & \cellcolor{lightgreen}accept & \cellcolor{lightgreen}accept & \cellcolor{lightgreen}accept & \cellcolor{lightgreen}accept & \cellcolor{lightgreen}accept & \cellcolor{lightgreen}accept & \cellcolor{lightgreen}accept \\ 
ModelB & \cellcolor{lightgreen}accept & \cellcolor{lightgreen}accept & \cellcolor{lightgreen}accept & \cellcolor{cream}check & \cellcolor{salmon}reject & \cellcolor{cream}check & - & - \\ 
ModelC & \cellcolor{lightgreen}accept & \cellcolor{lightgreen}accept & \cellcolor{lightgreen}accept & \cellcolor{lightgreen}accept & \cellcolor{lightgreen}accept & \cellcolor{lightgreen}accept & - & - \\ 
ModelD & \cellcolor{lightgreen}accept & \cellcolor{lightgreen}accept & \cellcolor{lightgreen}accept & \cellcolor{salmon}reject & \cellcolor{salmon}reject & \cellcolor{cream}check & - & - \\ 
ModelE & \cellcolor{lightgreen}accept & \cellcolor{lightgreen}accept & \cellcolor{lightgreen}accept & \cellcolor{lightgreen}accept & \cellcolor{lightgreen}accept & \cellcolor{lightgreen}accept & - & - \\ 
ModelF & \cellcolor{lightgreen}accept & \cellcolor{lightgreen}accept & \cellcolor{lightgreen}accept & \cellcolor{cream}check & \cellcolor{salmon}reject & \cellcolor{cream}check & - & - \\ 
ModelG & \cellcolor{lightgreen}accept & \cellcolor{lightgreen}accept & \cellcolor{lightgreen}accept & \cellcolor{lightgreen}accept & \cellcolor{lightgreen}accept & \cellcolor{lightgreen}accept & - & - \\ 
Sun01 & \cellcolor{lightgreen}accept & \cellcolor{lightgreen}accept & \cellcolor{lightgreen}accept & \cellcolor{lightgreen}accept & \cellcolor{lightgreen}accept & \cellcolor{lightgreen}accept & \cellcolor{lightgreen}accept & \cellcolor{cream}check \\ 
16CygA & \cellcolor{lightgreen}accept & \cellcolor{lightgreen}accept & \cellcolor{lightgreen}accept & \cellcolor{lightgreen}accept & \cellcolor{lightgreen}accept & \cellcolor{lightgreen}accept & - & - \\ 
16CygB & \cellcolor{lightgreen}accept & \cellcolor{lightgreen}accept & \cellcolor{lightgreen}accept & \cellcolor{lightgreen}accept & \cellcolor{lightgreen}accept & \cellcolor{lightgreen}accept & - & - \\ 
Kepler93 & \cellcolor{lightgreen}accept & \cellcolor{cream}check & \cellcolor{lightgreen}accept & \cellcolor{cream}check & \cellcolor{lightgreen}accept & \cellcolor{cream}check & \cellcolor{salmon}reject & \cellcolor{lightgreen}accept \\ 
AlphaCentA & \cellcolor{lightgreen}accept & \cellcolor{lightgreen}accept & \cellcolor{lightgreen}accept & \cellcolor{salmon}reject & \cellcolor{salmon}reject & \cellcolor{lightgreen}accept & \cellcolor{salmon}reject & \cellcolor{salmon}reject \\ 
AlphaCentAOv & \cellcolor{lightgreen}accept & \cellcolor{lightgreen}accept & \cellcolor{salmon}reject & \cellcolor{salmon}reject & \cellcolor{salmon}reject & \cellcolor{cream}check & \cellcolor{salmon}reject & \cellcolor{salmon}reject \\ 
AlphaCentB & \cellcolor{lightgreen}accept & \cellcolor{cream}check & \cellcolor{lightgreen}accept & \cellcolor{salmon}reject & \cellcolor{lightgreen}accept & \cellcolor{salmon}reject & \cellcolor{salmon}reject & \cellcolor{salmon}reject \\ 
AlphaCentBOv & \cellcolor{lightgreen}accept & \cellcolor{cream}check & \cellcolor{lightgreen}accept & \cellcolor{cream}check & \cellcolor{lightgreen}accept & \cellcolor{salmon}reject & \cellcolor{salmon}reject & \cellcolor{salmon}reject \\ 
Arthur & \cellcolor{lightgreen}accept & \cellcolor{lightgreen}accept & \cellcolor{lightgreen}accept & \cellcolor{lightgreen}accept & \cellcolor{lightgreen}accept & \cellcolor{lightgreen}accept & - & - \\ 
Baloo & \cellcolor{lightgreen}accept & \cellcolor{cream}check & \cellcolor{lightgreen}accept & \cellcolor{lightgreen}accept & \cellcolor{lightgreen}accept & \cellcolor{cream}check & - & - \\ 
Barney & \cellcolor{lightgreen}accept & \cellcolor{lightgreen}accept & \cellcolor{lightgreen}accept & \cellcolor{lightgreen}accept & \cellcolor{lightgreen}accept & \cellcolor{lightgreen}accept & - & - \\ 
Carlsberg & \cellcolor{lightgreen}accept & \cellcolor{lightgreen}accept & \cellcolor{lightgreen}accept & \cellcolor{lightgreen}accept & \cellcolor{lightgreen}accept & \cellcolor{lightgreen}accept & - & - \\ 
Doris & \cellcolor{lightgreen}accept & \cellcolor{lightgreen}accept & \cellcolor{lightgreen}accept & \cellcolor{lightgreen}accept & \cellcolor{lightgreen}accept & \cellcolor{cream}check & - & - \\ 
Dushera & \cellcolor{lightgreen}accept & \cellcolor{lightgreen}accept & \cellcolor{lightgreen}accept & \cellcolor{salmon}reject & \cellcolor{salmon}reject & \cellcolor{lightgreen}accept & - & - \\ 
Nunny & \cellcolor{lightgreen}accept & \cellcolor{lightgreen}accept & \cellcolor{lightgreen}accept & \cellcolor{lightgreen}accept & \cellcolor{lightgreen}accept & \cellcolor{lightgreen}accept & - & - \\ 
Pinocha & \cellcolor{lightgreen}accept & \cellcolor{lightgreen}accept & \cellcolor{lightgreen}accept & \cellcolor{lightgreen}accept & \cellcolor{lightgreen}accept & \cellcolor{cream}check & - & - \\ 
Punto & \cellcolor{lightgreen}accept & \cellcolor{lightgreen}accept & \cellcolor{lightgreen}accept & \cellcolor{lightgreen}accept & \cellcolor{lightgreen}accept & \cellcolor{lightgreen}accept & - & - \\ 
Saxo2 & \cellcolor{lightgreen}accept & \cellcolor{lightgreen}accept & \cellcolor{lightgreen}accept & \cellcolor{lightgreen}accept & \cellcolor{lightgreen}accept & \cellcolor{lightgreen}accept & - & - \\ 
Tinky & \cellcolor{lightgreen}accept & \cellcolor{lightgreen}accept & \cellcolor{lightgreen}accept & \cellcolor{lightgreen}accept & \cellcolor{salmon}reject & \cellcolor{cream}check & - & - \\ 
\hline 
\end{tabular}
{\par\small\justify\textbf{Notes.}  We considered four semi-empirical surface effect prescriptions: no surface corrections (no surf.), \citet{Ball&Gizon2014} two terms (BG14), \citet{Sonoi2015} two terms (TS15), and a sixth order polynomial (solar $6^{\mathrm{th}}$ order). \par}
\label{tab:tau_flags}
\end{table*}

In Table \ref{tab:tau_flags}, we show the assessment flags of the acoustic inversions carried out on our testing set. As mentioned in Sec. \ref{sec:results_rho_tau_inversions}, the quality behaviour of a mean density inversion and of an acoustic radius inversion is similar. We therefore invite the reader to refer to this section for the interpretation of the results.

\end{appendix}
\end{document}